\documentclass[Mena-Andrade2026,preprint,review,12pt]{elsarticle}

\usepackage{amssymb}
\usepackage{amsmath}

\usepackage{lineno}
\usepackage{subcaption} 

\journal{Journal of Manufacturing Processes}

\begin{document}
\begin{frontmatter}



\title{Manufacturability studies for the FCC-ee positron source target: determination of the minimum bending radius and ovalization in tantalum cooling tube elbows}

\author{Ramiro Mena-Andrade}
\ead{ramiro.francisco.mena.andrade@cern.ch}
\author{Mickaell Crouvizier} %
\author{Jean-Philippe Rigaud}%
\author{Thibaut Coiffet}
\author{Antonio Perillo-Marcone}
\ead{antonio.perillo-marcone@cern.ch}
\affiliation{organization={CERN},
            addressline={1211 Geneva 23}, 
            country={Switzerland}}


\begin{abstract}

Beam intercepting devices rely on cooling systems to effectively dissipate the thermal energy generated during the impact of a high-energy beam. Regardless of the device’s size, integrating the cooling system is a complex task, particularly when the resulting device is only a few centimetres in size, as is the case with the positron source target for the Future Circular Collider (FCC-ee) at CERN, where the current design consists of a tungsten core with two embedded tantalum cooling tubes.

Due to the reduced dimensions of the chosen tantalum tubes (OD6.35xID4.35 mm), the selected manufacturing method is compression bending. The present study develops and evaluates a numerical model to manufacture the required elbow. The methodology is divided in four steps: \emph{i)} minium allowable bending radius calculation, \emph{ii)} material constitutive law validation, \emph{iii)} prediction of the resulting distortion due to ovalization and \emph{iv)} experimental validation via (non) destructive methods.

The results indicate that a minimum bending radius of 10 mm is suitable for manufacturing the elbow. The distortion caused by ovalization is within ±0.5 mm, resulting in an important deviation respect to the nominal geometry. The numerical model was successfully validated experimentally. The micrographies performed in the cross-section of the tantalum tube before and after plastic bending confirm the integrity of the elbow. Additionally, an empirical expression is proposed to estimate the yield stress of pure tantalum based on Vickers hardness measurements.

The proposed numerical model is capable to predict the ovalization along the resulting elbow, offering a viable alternative to define the cooling tube geometry. This study provides a methodology to determine the minimum bending radius for thick walled tubes to be used with compression bending and can be applied for the cooling system design of other high-performance devices.

\end{abstract}



\begin{keyword}
Future Circular Collider (FCC) \sep Positron source target \sep cooling tubes \sep tantalum \sep plastic bending \sep compression bending \sep ovalization



\end{keyword}

\end{frontmatter}



\section{Introduction}\label{sec:Intro}

The Future Circular Collider (FCC) is the next accelerator facility foreseen to continue the endeavor of the Large Hadron Collider (LHC) at CERN after the High Luminosity HiLumi-LHC era. The FCC International Collaboration released a detailed feasibility study report to the Particle Physics' research community in 2025. As a first stage, FCC-ee will use a high-intensity electron beam to produce positrons by hitting a fixed target due to the photons generated first by Bremsstrahlung, followed by the electron-positron pair production mechanism \cite{Benedikt2025, Hubbell2006}.

\begin{figure}[!htb]
\includegraphics[width=\textwidth]{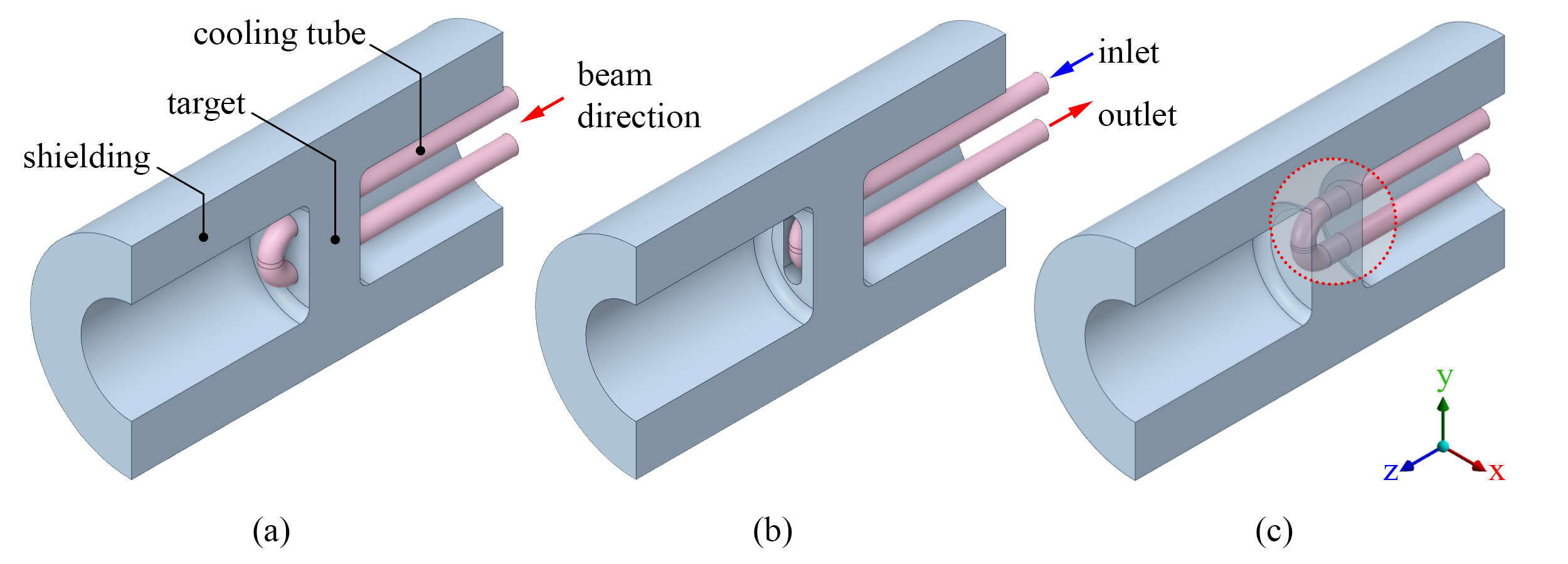}
\caption{FCC-ee positron source target geometry: (a) external, (b) tangential and c) embedded cooling tubes options. Note: only one half of the geometry is represented.}
\label{fig:model_geometry_manufacturing}
\end{figure}

From the thermo-mechanical point of view, the target, as a beam intercepting device (BID), needs to properly dissipate the power deposited by the incoming high-energy electron beam. Based on the beam parameters reported on \cite{Benedikt2025} (for a 2.86 GeV electron beam, with a spot size $\sigma_x$ of 1 mm, a bunch charge of 3.8 nC, with 4 bunches at a frequency of 100 Hz, from the resulting 4.3 kW of the e$^-$ beam, a target made of pure tungsten with a thickness of 15 mm will receive around 1 kW of deposited power), the target must include a proper cooling system to keep the temperature and the resulting thermal stresses within the safety limits of the material.

Currently, tungsten (W) is the candidate material to be used for the FCC-ee positron source target. This selection is based on the high-Z number (74), high density (19300 kg/m$^3$), high melting point (3.422 $^{\circ}$C) and remarkable mechanical properties at high temperatures \cite{Lassner1999}. As a drawback, tungsten cannot be in direct contact with an active cooling media (e.g. air, H$_2$O, CO$_2$) due to generation of tungsten oxides (WO$_2$ and WO$_3$)\cite{Cardillo1977}. At the temperature range of 700-800 $^{\circ}$C, there is a potential risk of WO$_3$ vaporization  when in contact with steam \cite{Greene2001}. To mitigate this problem, one possible solution is to isolate tungsten from the cooling fluid by using a barrier material as tantalum. 

Tantalum (Ta) is a corrosion resistant element, highly used in the chemical industry \cite{Cardone1995}. In high-energy physics, tantalum is used as a cladding for tungsten core targets (e.g KENS \cite{Kawai2001}, and ISIS \cite{Wilcox2018}), were a continuous bonding between dissimilar materials is obtained by diffusion bonding via Hot Isostatic Pressing (HIP) technology \cite{Busom2020}. In addition, tantalum is ductile at room temperature and this property allows to consider geometries as complex as  a serpentine or simpler as an elbow for the cooling circuit. 

Figure \ref{fig:model_geometry_manufacturing} shows a half model of the FCC-ee positron source target, due to its symmetry in the x-axis, where a 180$^\circ$ elbow was chosen for the cooling system, so that the input and output points are located on the same side of the device. During the design phase, multiple configurations were considered. Starting from option $(a)$, where the tubes traverse the target and the elbow is placed externally, just after the exit surface of the beam. From the physics performance, it is preferred to not have a material in contact with the produced particles after the exit surface, therefore, this option was rejected. Next, option $(b)$, presents a cooling tube placed tangentially to the exit surface of the device. To this end, a complex grove would be needed to host the tube. However, this option was discarded, due to the incompatibility to use HIP to join the tube to the target because of the absence of material to support the elbow with the target during the diffusion bonding process. Finally, option $(c)$ presents an embedded tantalum cooling tube, placed inside of the target. Currently, option $(c)$ is the preferred configuration for the target cooling system.

\begin{figure}[!ht]
\centering
\includegraphics[width=0.5\textwidth]{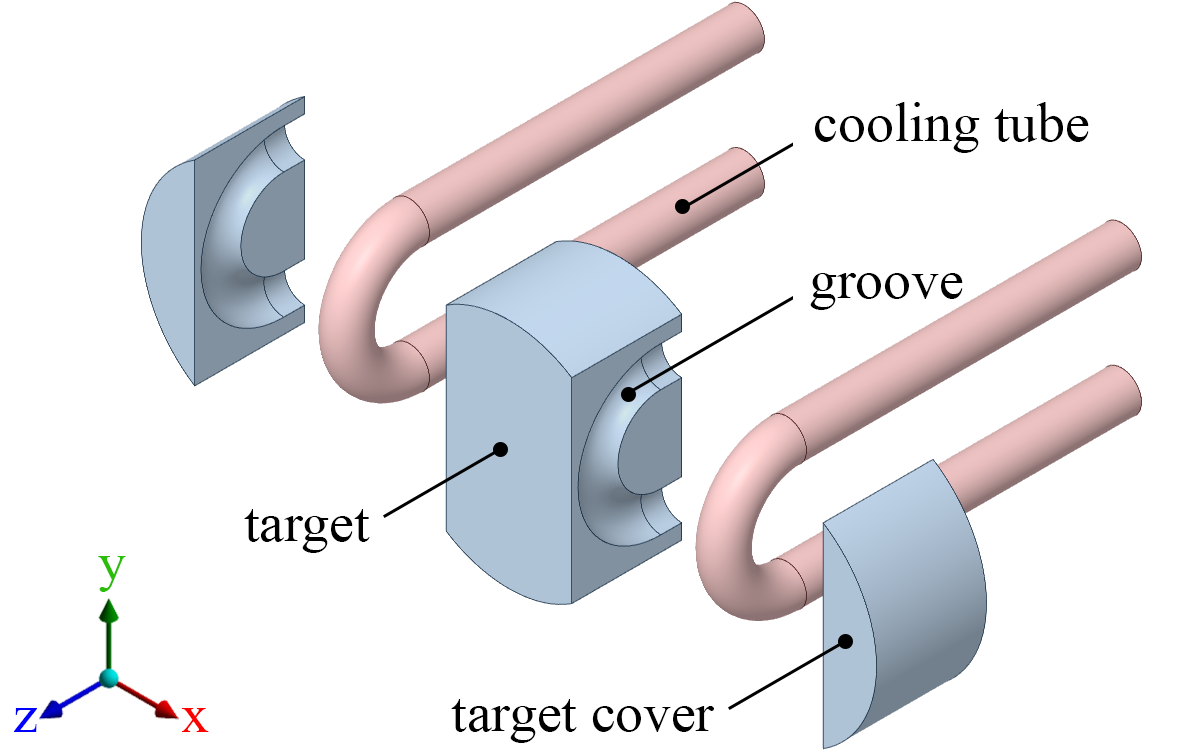}
\caption{Exploded view of the FCC-ee positron source target (without shielding).}
\label{fig:capsule}
\end{figure}

Due to the symmetric design, the placement of the $n$ cooling tubes inside of the target, requires to slice the device in $n+1$ parts and machine $2n$ half grooves with the real form of the elbow, as shown in Figure \ref{fig:capsule}. As the resulting geometry of the elbow is different from its nominal shape, due to the phenomenon referred as \emph{ovalization} (see Section \ref{sec:Compression_bending}), it is mandatory to know the real shape in advance. Currently, this issue is solved by scanning the real tubes and subtracting the resulting geometry in the CAD model. However, before performing this operation, and due to the space limitations of the target, it is required to determine the minimum bending radius to obtain an elbow without breaking the tubes.  

The main motivation of the present work is to include the manufacturing constrains in the FCC-ee positron source design. To this end, a numerical model of compression bending is developed to study the deformed shape of the elbow. By finding the minimum bending radius, the geometry of the cooling system can be defined. In addition, the model can provide valuable insights in terms of resulting residual stresses developed during the process, giving hints and guidelines to be used in the further development of the target.  

This document is organized as follows. Section 2 introduces the manufacturing operation of compression bending, explaining its main characteristics, advantages and limitations. Here, the criteria used to select the minimum bending radius R$_b$ are explained. Then, Section 3 describes the numerical and experimental methods used to validate the selected bending radius. The results in terms of stresses, displacements, microstructure and hardness measurements along the elbow are presented in Section 4. A discussion is followed in Section 5 where the potential advantages and limitations of the present study are highlighted. Finally, in Section 6, the concluding remarks are summarized.

\section{Compression bending}
\label{sec:Compression_bending}

Compression bending is the used manufacturing operation to conform the tube elbow for the FCC-ee positron source's cooling system. The choice was driven by the reduced dimensions of the selected commercial tantalum seamless tube: OD/ID 6.35/4.35 mm, where OD and ID stands for outside and inside diameter, respectively. For this case, the use of rotary-draw bending is not recommended\footnote{For tubes of larger diameters with wall thickness defined as thin or below, rotary-draw bending process is recommended due to the presence of an articulated mandrel that provides internal support  to the hollow tube section while deformed plastically. As a consequence, a higher precision degree is obtained when compared to compression bending. As a result, rotary-draw bending method is widely used for the construction of cooling systems for Beam Intercepting Devices (BID) at CERN as the Target Internal Dump Vertical (TIDV) 5 \cite{Romero2024} or the Large Hadron Collider (LHC) collimators \cite{Bertarelli2004}.}. Below, a brief overview of compression bending process, followed by the estimation of the minimum bending radius are provided.

\subsection{Process overview}

\begin{figure}[!ht]
     \centering
     \begin{subfigure}[b]{0.325\textwidth}
         \centering
         \includegraphics[width=\textwidth]{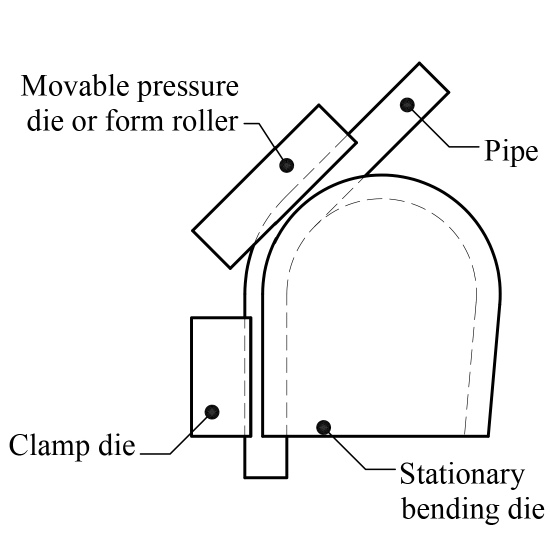}
         \caption{}
         \label{fig:CB_scheme}
     \end{subfigure}
     \hfill
     \begin{subfigure}[b]{0.65\textwidth}
         \centering
         \includegraphics[width=\textwidth]{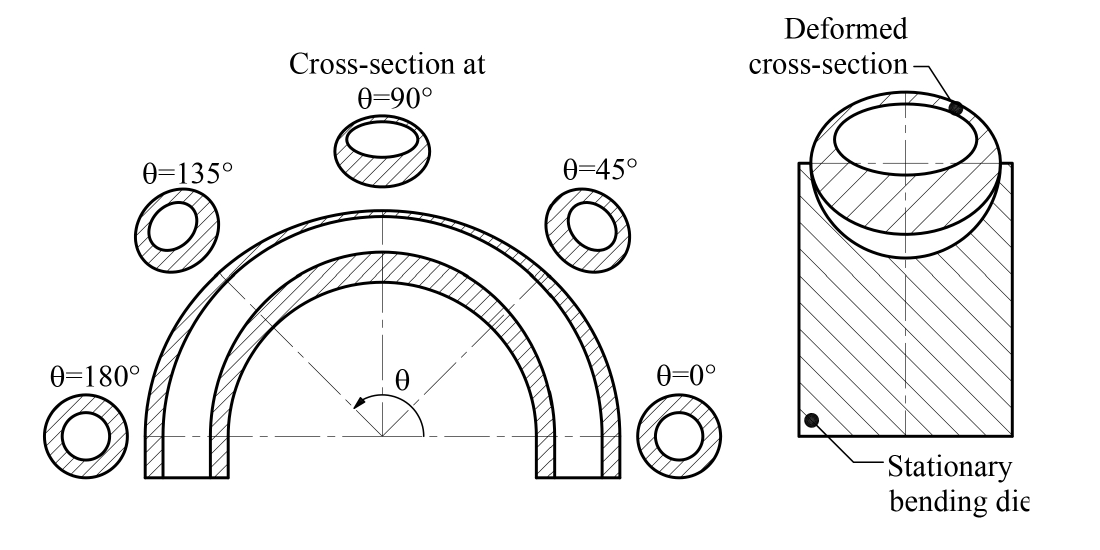}
         \caption{}
         \label{fig:CB_ovlization}
     \end{subfigure}
     \hfill
     \caption{Compression bending: (a) Process schematic representation and (b) Evolution of the cross-section ovalization for a 180$^\circ$ elbow. Adapted from \cite{Olofson1961, Pan1995}}
        \label{fig:Compression_bending}
\end{figure}

Compression bending is a cold-working manual process suitable for ductile metals as copper, stainless steel and tantalum. A graphical representation of this manufacturing process is depicted in Figure \ref{fig:CB_scheme}. The operation is composed by a clamp die in charge of providing support to the part during the loading and unloading phase. Then, the tool is divided in two parts: \emph{i)} a stationary die with a predefined bending radius $R_b$ and \emph{ii)} a movable pressure die that progressively deforms the tube, giving the final shape after rotating a defined bending angle $\theta$, where a maximum of 180$^\circ$ can be obtained. However, during the unloading phase, springback takes places. Therefore, overbending is required to compensate the elastic recovery to achieve the desired final bending angle. As the forming dies are in direct contact with the tube, proper lubrication of the tooling is needed to reduce friction effects. For a complete review of plastic bending methods in general, the interested reader is refereed to Olofson \cite{Olofson1961}.

During plastic deformation, the wall portion subjected to compression undergoes an increase in thickness while the outer fibers working under tension become thinner. As a result, there is a progressive deviation of neutral axis, the deformed cross section experiences a shrinking rate producing a phenomenon defined as \emph{ovalization} and the final geometry differs from the nominal cross-section \cite{Pan1995}, as represented schematically in Figure \ref{fig:CB_ovlization}. This phenomenon is unavoidable, it is the responsible that the moment of inertia is not constant along the deformed cross-section and it depends on the tube geometry, the the material properties and the selected bending radius. 

Important research efforts has been done to predict the cross-section deformation after plastic bending. Pan et al., \cite{Pan1995} proposed an analytical model to predict the ovalization by considering the wall thickness variation. To this end, the displacement field was approximated by a trigonometric form capable to capture the neutral axis shift during bending. The final goal of the study was to improve the springback angle in a real-time control during the process. Also Tang \cite{Tang2000} used the plastic-deformation theory to develop a group of seven analytical expressions to explain tube-bending phenomena: $i)$ stresses in the bend; $ii)$ wall thickness change; $iii)$ shrinking rate at the tube section; $iv)$ deviation of neutral axis; $v)$ feed preparation length of the bend; $vi)$ bending moment, and $vii)$ flattening. As a limitation, the model was developed under the assumption of thin-wall thickness, where the radial stress $\sigma_r$ is neglected, leaving the case of thick walled tubes unattended.

Al-Qureshi \cite{AlQureshi1999} developed an analytical method for studying the elastic-plastic behavior of tube bending for the case of a perfect plastic material. Then, El-Megharbel et al, \cite{El-Megharbel2008} generalized the previos case by considering a material with strain hardening in the form of a power law $\sigma = C\varepsilon^n$. In parallel, Murata et al, \cite{Murata2008} explored the influence of the strain hardening exponent $n$, by keeping the tube geometry and bending radius fixed. It was reported that $n$ plays a minor roll on springback, thickness strain distribution,  and flatness ratio. Following this line, Paulsen et al, \cite{Paulsen2003} developed an analytical model to predict the ovalization in inelastic bending. In their work, the resulting radial distortion $\delta$ is expressed as a function of the following parameters (in order of importance): $\delta=f(D,R_b,t,n,\Gamma,b)$, diameter of the tube $D$, bending radius $R_b$, wall thickness $t$, strain hardening exponent \emph{n}, Gamma function $\Gamma$, and the plastic offset strain $b$. For the case of a material without strain hardening ($n$=0), the last three parameters can be dropped and the expression can be written as $\delta = \frac{3}{128}\frac{D^5}{t^2R_b^2}$. 

Tronvoll et al, \cite{Tronvoll2023} compared (roller based) compression bending against rotary-draw-bending for a 6060-T4 aluminum alloy tube of $OD$ 57 mm with a wall thickness $t$ 3 mm and bending radius of 222 mm in terms of performance (e.g. springback angle, cross-sectional distortion and wall thinning/thickening) and it was found that altough rotary draw bending is the preferred solution in the industry due to its thigh-tolerance, compression bending can be an alternative option that generates less pronounced springback with a small penalty in terms of dimensional accuracy presented as an increased cross-section deformation.

Once the bending process and material are selected, the only remaining parameter to define is the bending radius. While the use of a value above the minimum radius should not generate any constrain, except the use of more material, a value below that threshold will produce the failure of the tube by fracture in the tensile side or by buckling in the compressive side (for thin-walled tubes) \cite{Munz1982}. Therefore, the proper selection of the minimum bending radius is a key parameter that is explained below.

\subsection{Miminum bending radius estimation}

\begin{figure}[!htb]
    \centering
     \begin{subfigure}[b]{0.32\textwidth}
         \centering
         \includegraphics[width=\textwidth]{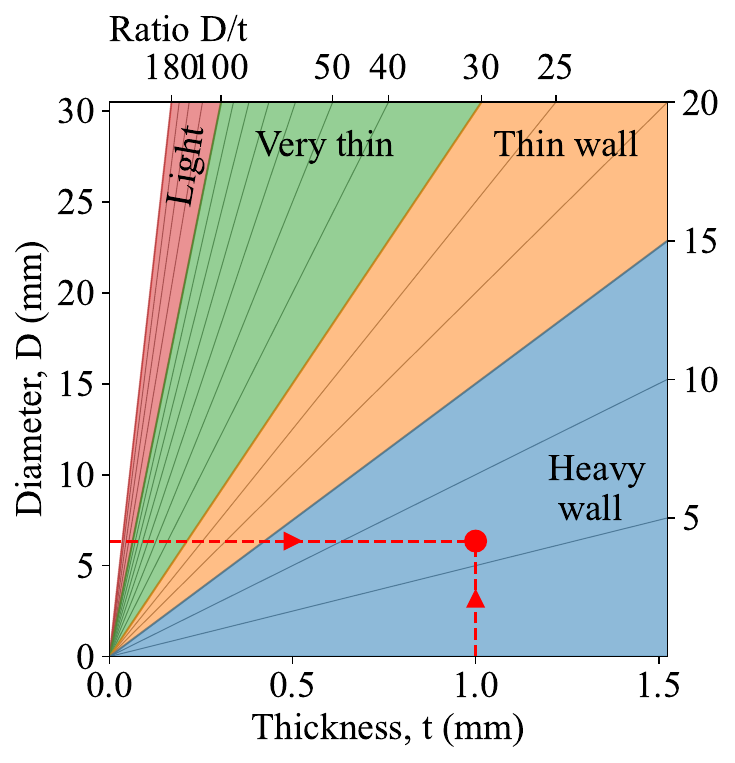}
         \caption{}
         \label{fig:Nomograph}
     \end{subfigure}
     \hfill
     \centering
     \begin{subfigure}[b]{0.32\textwidth}
         \centering
         \includegraphics[width=\textwidth]{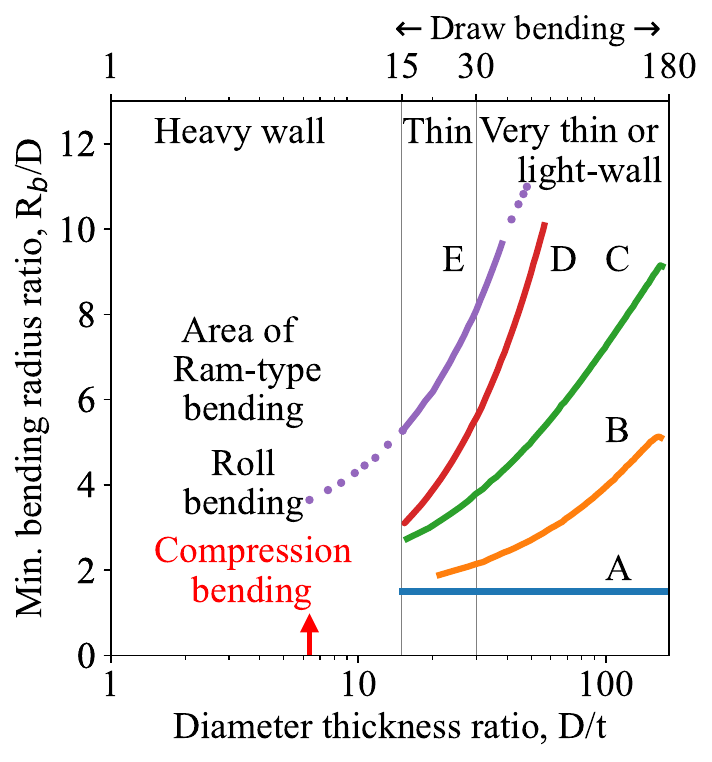}
         \caption{}
         \label{fig:Min_Radius_criteria}
     \end{subfigure}
     \hfill
     \vspace{0.1pt}
     \begin{subfigure}[b]{0.32\textwidth}
         \centering
         \includegraphics[width=\textwidth]{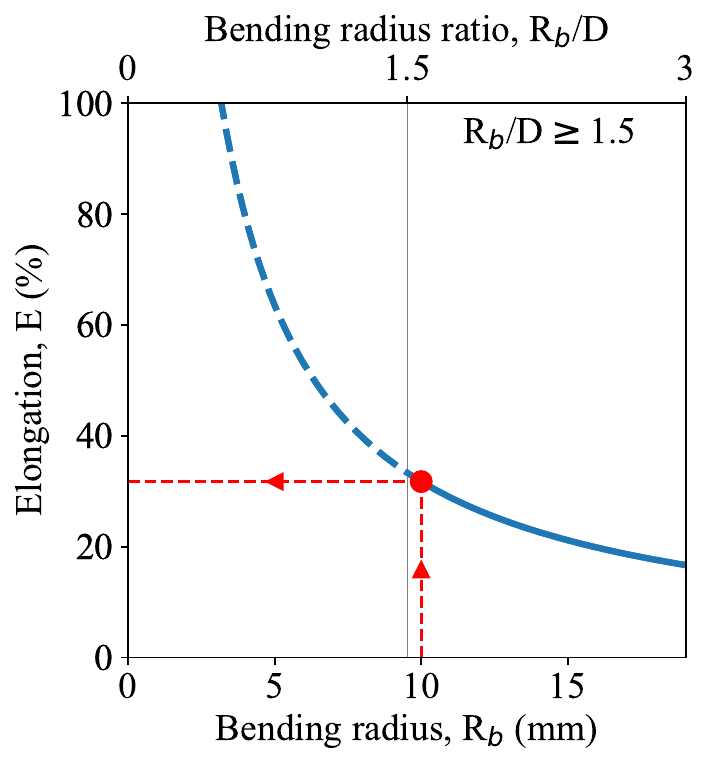}
         \caption{}
         \label{fig:Min_Radius}
     \end{subfigure}
     \hfill
     \caption{Minimum bending radius estimation process: (a) Step 1: Definition of the D/t ratio on the D/t nomograph, (b) Step 2: Selection of the minimum bending radius criteria and (c) Step 3: Elongation estimation based on the selected bending radius. Adapted from \cite{Miller2003,Olofson1961,Springborn1966}}
        \label{fig:Minimum_bendingi_radius_estimation}
\end{figure}

To estimate the minimum bending radius, the selection process was performed in three steps:
\begin{itemize}
\item \emph{Step 1}: Definition of the diameter to thickness ratio $D/t$ using the $D/t$ nomograph depicted in Figure \ref{fig:Nomograph}. For our case, the tantalum tube presents a $D/t$=6.35. As the resulting $D/t\leq$15, it is considered as a heavy wall (or thick walled) tube. 
\item \emph{Step 2}: Using the obtained $D/t$ as an input for Figure \ref{fig:Min_Radius_criteria}, the plot provides the minimum bending radius from various bending processes. Even though no specific criteria was found for the case of compression bending of thick walled tubes, the most restrictive recommendation for rotary draw-bending with multi ball-mandrel (case A in Figure \ref{fig:Min_Radius_criteria}) was used as a guideline, where the minimum bending radius ratio $R_b/D\geq1.5$ is provided.
\item \emph{Step 3:} Calculation of cold-working elongation $E$ $(\%)$ by using the general expression $E=\frac{D}{2R_b}$, shown in Figure \ref{fig:Min_Radius}. The plot is divided in two regions by including the previously defined minimum bending radius ratio $R_b/D\geq1.5$ as a pass-fail filter, leaving the zone of interest (continuous line) on the right. The selected minimum bending radius is 10 mm (after rounding to the next digit), giving an estimated elongation of 31.75 \% (depicted in red). 
\end{itemize}

\section{Method}
\label{sec:Model_setup}

This section contains the three phases performed to validate the tantalum elbow tube manufacturing process. Starting from a previous available tensile test, the resulting true stress-strain curve was reproduced numerically by using a material model as a calibration step. Then, once the material model was able to replicate the tantalum behavior in the plastic regime, the compression bending operation was numerically simulated and finally, the real manufacturing operation was performed on the workshop. The detailed setup used at each step is described below.

\subsection{Material model}
\begin{figure}[!htb]
    \centering
     \begin{subfigure}[b]{0.32\textwidth}
         \centering
         \includegraphics[width=0.5\textwidth]{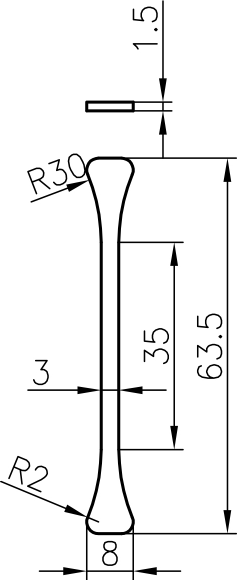}
         \caption{}
         \label{fig:tensile_test_sample}
     \end{subfigure}
     \hfill
    \centering
    \begin{subfigure}[b]{0.32\textwidth}
         \centering
         \includegraphics[width=\textwidth]{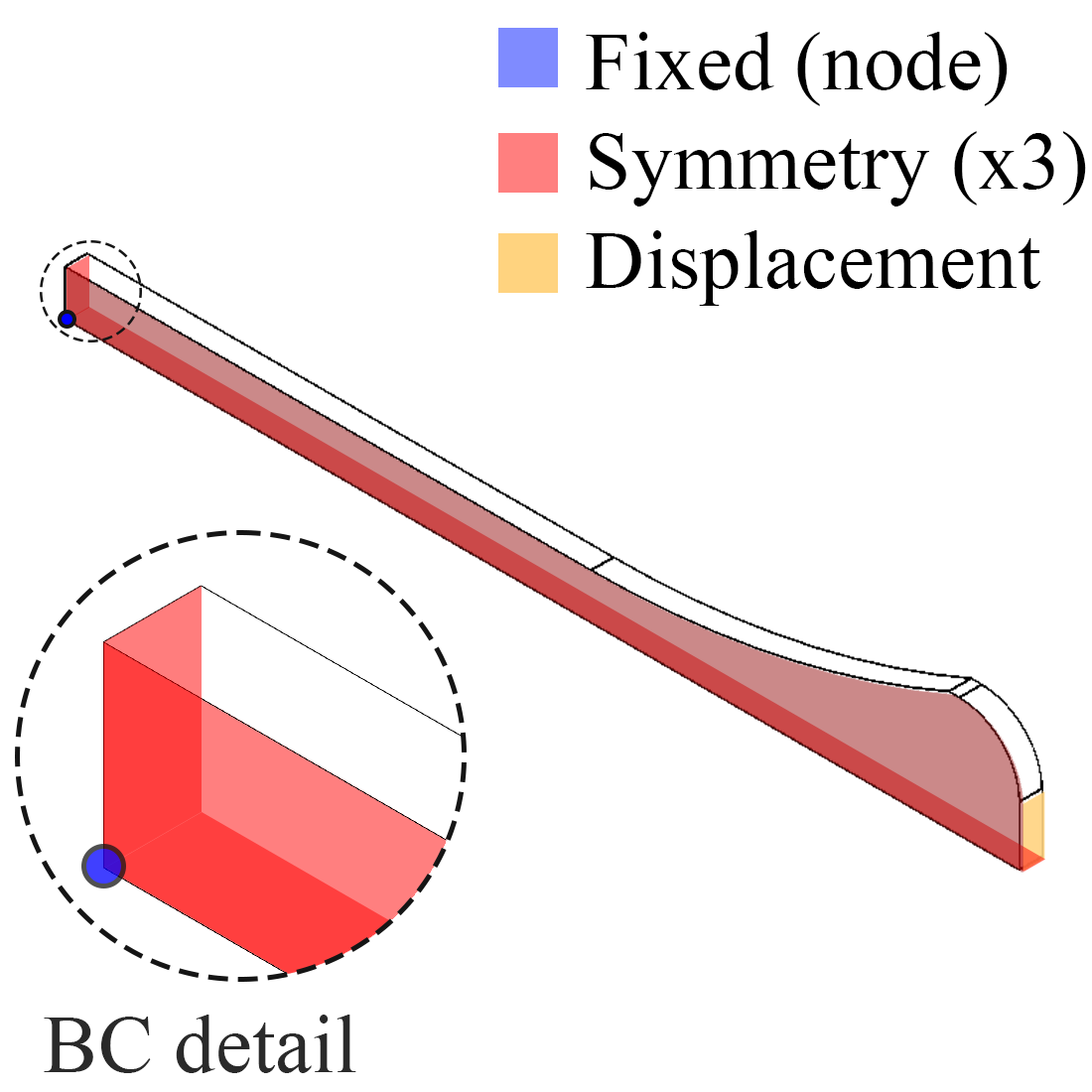}
         \caption{}
         \label{fig:tensile_test_BC}
     \end{subfigure}
     \hfill
     \centering
     \begin{subfigure}[b]{0.32\textwidth}
         \centering
         \includegraphics[width=\textwidth]{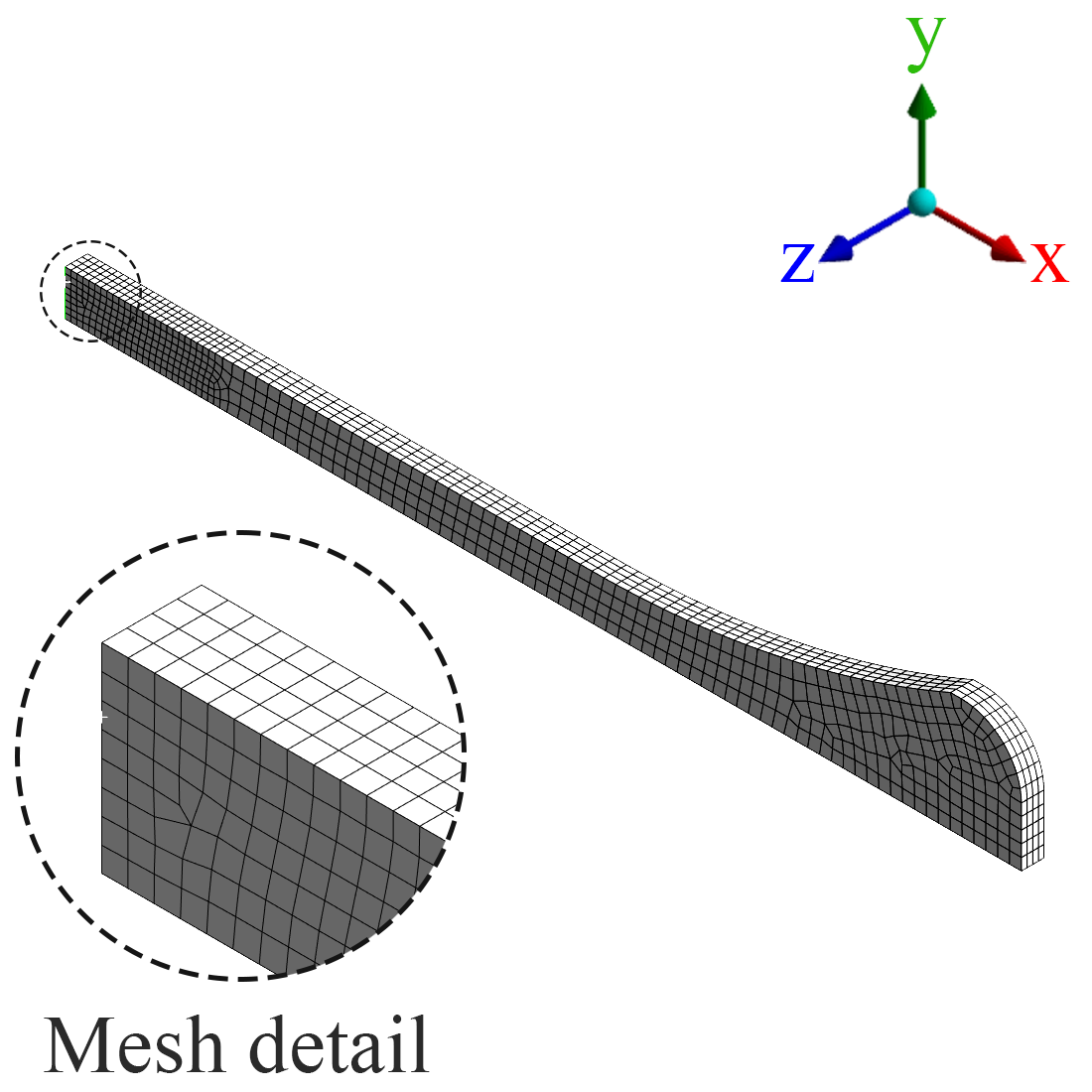}
         \caption{}
         \label{fig:tensile_test_mesh}
     \end{subfigure}
     \hfill
     \caption{Ta tensile test: (a) sample geometry, (b) boundary conditions, and (c) mesh.}
        \label{fig:tensile_test}
\end{figure}

The experimental data for the tantalum tensile test was taken from a previous test campaign done at CERN \cite{Fraunhofer2020} for the Beam Dump Facility (BDF) target \cite{LopezSola2019}. The geometry of the dog-bone specimen is depicted in Figure \ref{fig:tensile_test_sample}. Taking advantage of the symmetry, only a 1/8 sector was modelled, and the node at the origin was fixed to avoid solid rigid motion of the model. To mimic the tensile test before failure, a fixed displacement in the x-axis direction of 5.7 mm was imposed in the right surface of the sample. This displacement corresponds to 11.4 mm and it is lower to the 14 mm registered during the test. The discrepancy between the test and the simulation is due to the material model does not include a damage law. The applied boundary conditions are graphically represented in Figure \ref{fig:tensile_test_BC}.  For the mesh, a general element size of 0.4 mm was chosen with a refinement zone of 5 mm around the origin using an element size of 0.2 mm. The geometry was meshed with hexahedron elements type Hex20, with a total of 11393 nodes and 2096 elements, as depicted on Figure \ref{fig:tensile_test_mesh}. The model was implemented in the commercial Finite Element software Ansys 2022 R2 \cite{Ansys2022}.

From the material point of view, the selected constitutive model for tantalum is a simplified version of the Khan and Liang (KL) model \cite{Khan1999}. While the full KL model proposes a thermo-visco-plastic response $\sigma=f(\varepsilon,\dot{\varepsilon},T)$, the used simplified version was obtained as the tensile test was performed at room temperature, therefore, the thermal dependence term is equal to unity, resulting in a visco-plastic constitutive law $\sigma=f(\varepsilon,\dot{\varepsilon})$, as formulated below:
\begin{eqnarray}
    \sigma&=&\left[A+B\left(1-\frac{ln\dot{\varepsilon}}{ln D_0^p}\right)^{n_1}\varepsilon^{n_0}\right]\dot{\varepsilon}^C
\label{eq:KLmodel}
\end{eqnarray}

where $\sigma, \varepsilon$ and $\dot{\varepsilon}$ are Von Mises equivalent stress, strain and strain rate respectively. The simplified model is defined by five constants $A,B,C,n_0$ and $n_1$. Their respective values are summarized in Table \ref{tab:material_constants_KL1999}. To implement the material model, the reported value for constant $D_0^p$ is 10$^6$ s$^{-1}$. 
The tensile test was performed at a strain rate $\dot{\varepsilon}$,  of 9$\times$10$^{-4}$ s$^{-1}$. Regarding the elastic properties, a Young Modulus $E$ and a Poisson's ratio $\nu$ of values 182.75 GPa and 0.2775 were used, respectively. 

\begin{table}[!htb]
\caption{Material constants for the simplified KL-constitutive model \cite{Khan1999}}
\label{tab:material_constants_KL1999}
\centering
\begin{tabular}{ ccccccc } 
 \hline
 \textbf{Material} & \textbf{A [MPa]} & \textbf{B [MPa]} & \textbf{n$_0$} & \textbf{n$_1$} &\textbf{C}\\
 \hline
 Ta & 318.47 & 153.2 &0.6088 &3.1547&0.0759\\
 \hline
\end{tabular}
\end{table}

\subsection{Numerical model}
\begin{figure}[!htb]
    \centering
     \begin{subfigure}[b]{0.31\textwidth}
         \centering
         \includegraphics[width=\textwidth]{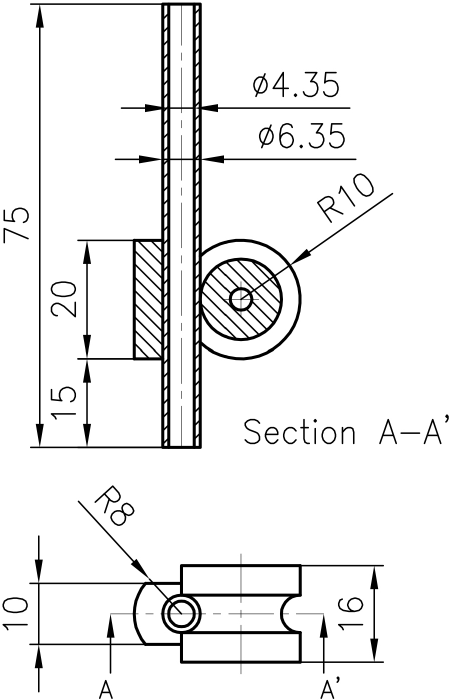}
         \caption{}
         \label{fig:plastic_bending_geometry}
     \end{subfigure}
     \hfill
    \centering
     \begin{subfigure}[b]{0.31\textwidth}
         \centering
         \includegraphics[width=\textwidth]{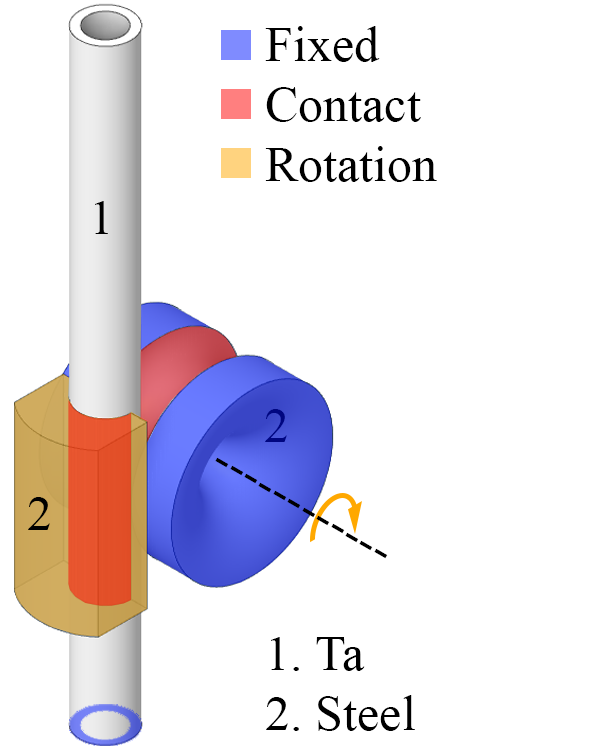}
         \caption{}
         \label{fig:plastic_bending_BC}
     \end{subfigure}
     \hfill
     \centering
     \begin{subfigure}[b]{0.31\textwidth}
         \centering
         \includegraphics[width=\textwidth]{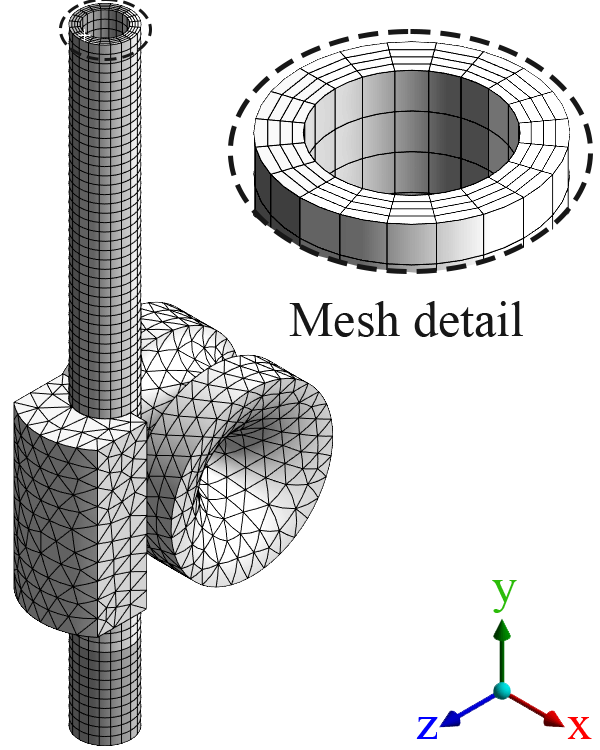}
         \caption{}
         \label{fig:plastic_bending_mesh}
     \end{subfigure}
     \hfill
     \caption{Compression bending numerical model: (a) geometry dimensions, (b) boundary conditions and material distribution, and (c) mesh.}
        \label{fig:plastic_bending_model}
\end{figure}

Once the bending radius $R_b$ of 10 mm was defined, the plastic bending operation was simulated numerically. Figure \ref{fig:plastic_bending_geometry} shows the geometry dimensions and the model setup. The applied boundary conditions are depicted in Figure \ref{fig:plastic_bending_BC}, where the bottom part of the tantalum tube and the stationary die (depicted in blue) were fixed. The movable die (in orange) was imposed a $\pm$180$^\circ$ clock-wise/counter-clockwise rotation around the x-axis to simulate the loading and unloading phase in two consecutive steps. Finally, the contact between both tools and the external surface of the tantalum tube was set up as friction-less as a simplification hypothesis. Note that for representation purposes, the tantalum tube contact surface was not coloured in red. 

Although the geometry presents symmetry around the x-axis, the full tube was simulated. While both dies were meshed with tetrahedron elements type Tet10 and an element size of 2 mm, the tube was divided in 4 semicircular sectors of 90$^\circ$. Each sector was segmented with 5 elements around the internal and external perimeters, so that the use of hexahedron elements type Hex20 was possible. For the tube, a general mesh size of 1 mm was chosen and 4 divisions through the thickness were imposed. As a result, the model presented a total of 37191 nodes and 11003 elements, as shown in the mesh depicted Figure \ref{fig:plastic_bending_mesh}. 

Regarding the materials, the tantalum material model described before was implemented and structural steel was used for the tools ($E$=193 GPa, $\nu$=0.31, $\sigma_y$=250 MPa). The model was implemented in the commercial Finite Element software Ansys 2022 R2 \cite{Ansys2022}.

\begin{figure}[!htb]
    \centering
     \begin{subfigure}[b]{0.31\textwidth}
         \centering
         \includegraphics[width=\textwidth]{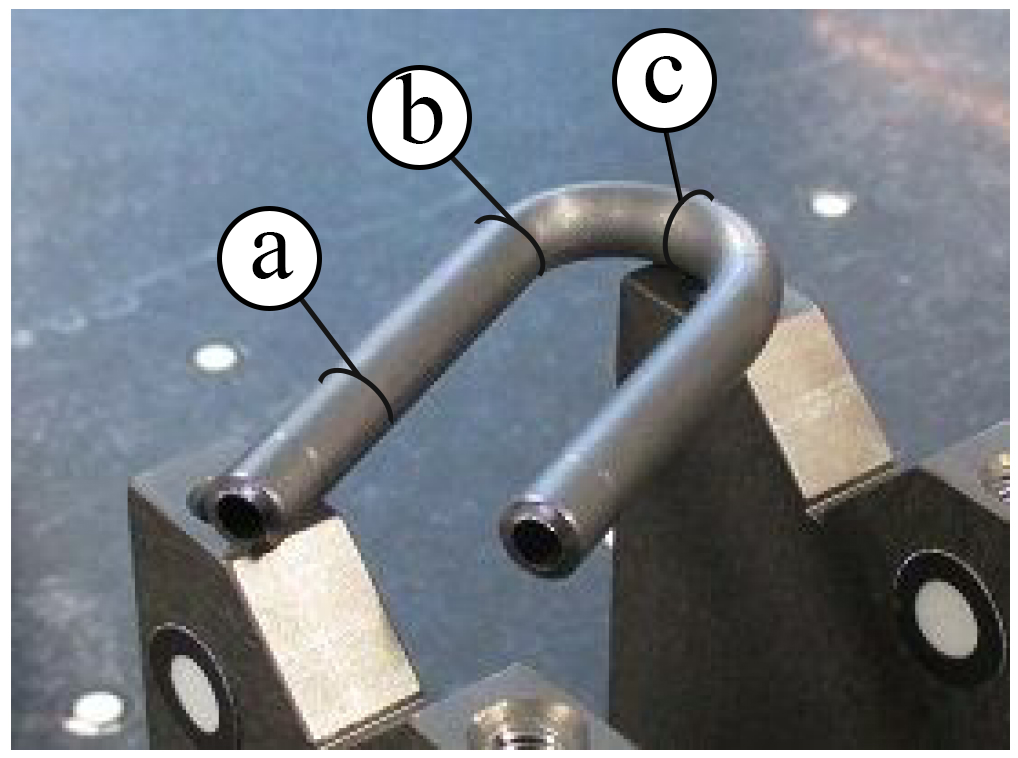}
         \caption{}
         \label{fig:metrology_test}
     \end{subfigure}
     \hfill
    \centering
     \begin{subfigure}[b]{0.62\textwidth}
         \centering
         \includegraphics[width=\textwidth]{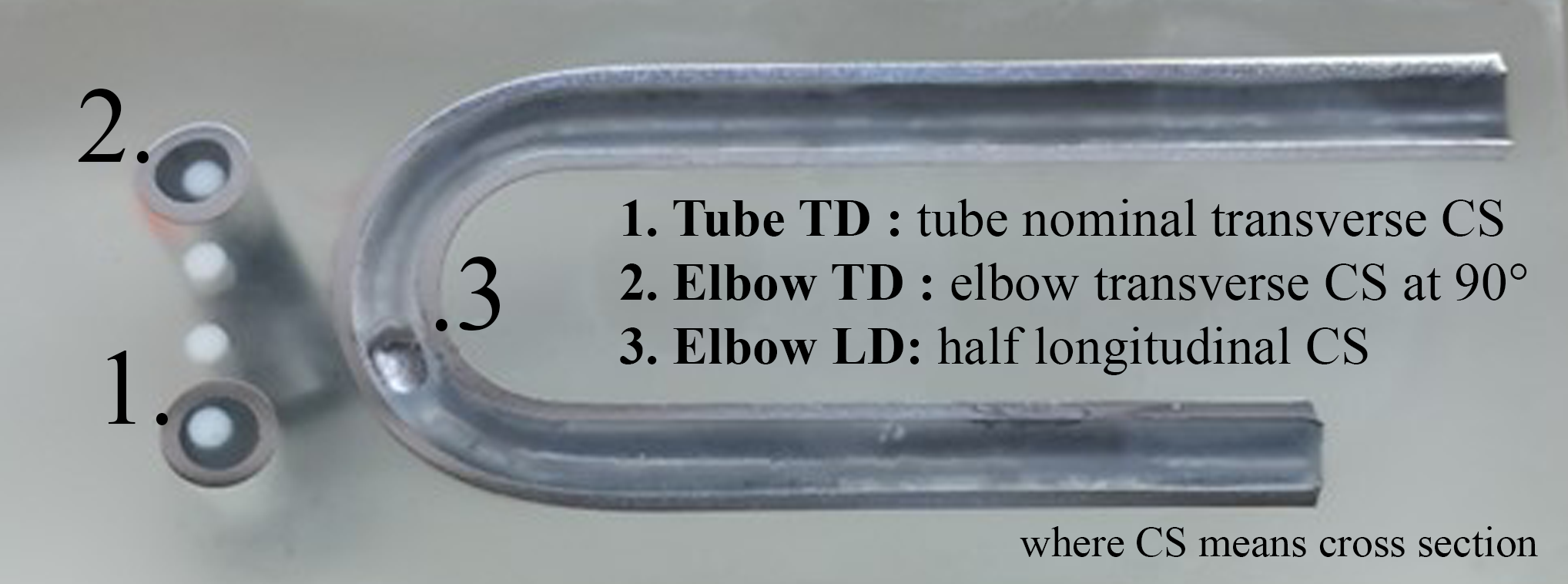}
         \caption{}
         \label{fig:micrograph_test}
     \end{subfigure}
     \hfill
     \caption{Experimental campaign tests setup: (a) metrology sample indicating the three locations of the perimeter measurements, and (b) metallographic specimen.}
        \label{fig:experimental_tests}
\end{figure}

\subsection{Experimental campaign}
Three elbows were manufactured in the Mechanical and Materials Engineering (MME) workshop at CERN. One sample was measured in the metrology laboratory to determine the distortion along the elbow while the other two samples were cut in different locations to perform micrographs and hardness tests. Each test setup is described below.

\subsubsection*{Metrology test}
One sample was subjected to a full 3D-scanner to obtain the distorted geometry envelop. In addition, the external perimeter of the tube was measured in three different locations: \emph{a)} below, \emph{b)} beginning, and \emph{c)} half of the elbow, respectively, as shown in Figure \ref{fig:metrology_test}. The goal was to quantify the distortion respect to the nominal geometry. For this test, the following equipment was used: 

\begin{itemize}
\item {Creaform HandsyScan, with a machine uncertainty of 20 $\mu$m + 40 $\mu$m/m.}
\end{itemize}

\subsubsection*{Micrography and hardness tests}
One longitudinal and two transverse cross sections were prepared following a dedicated grinding and polishing procedure. Etchant $\#$66 of ASTM E407 was used to reveal the tantalum microstructure. The resulting sample is shown in Figure \ref{fig:micrograph_test}. For each test, the following equipment was employed:
\begin{itemize}
    \item \emph{Micrography:} Zeiss AXIO Imager Z2.m optical microscope under bright field illumination. Used magnifications in the range of x50-500.
\item \emph{Hardness:} Wolpert Wilson 402 MVD indenter equipped with a Vicker tip with a load of 100 g. Five measurements per area of interest, at mid-thickness were taken.
\end{itemize}

\section{Results}
\label{sec:Results}
\subsection{Material model}
\label{ssec:Material_model}
\begin{figure}[!b]
    \centering
     \begin{subfigure}[b]{0.32\textwidth}
         \centering
         \includegraphics[width=\textwidth]{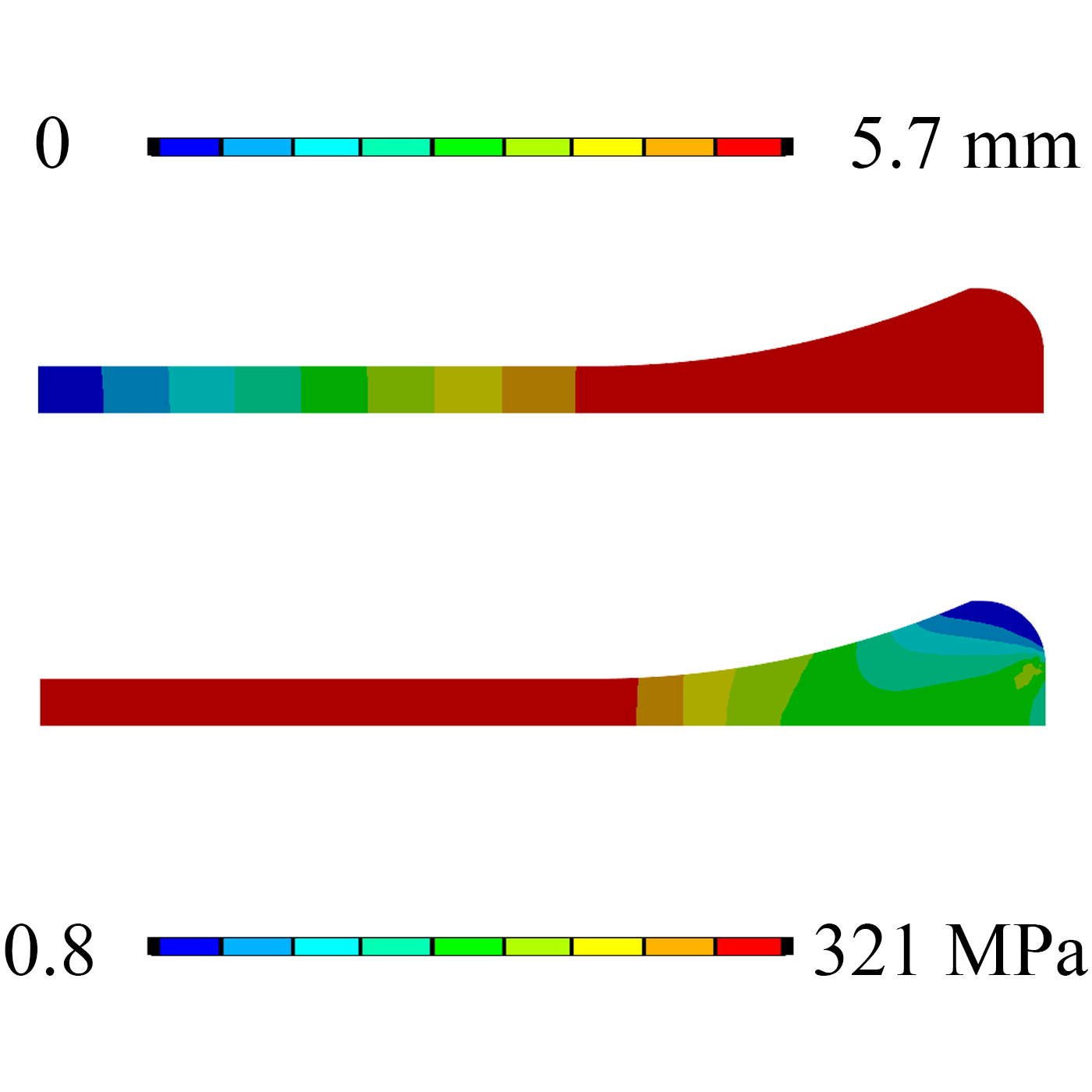}
         \caption{}
         \label{fig:mat_model_tensile}
     \end{subfigure}
     \hfill
     \centering
     \begin{subfigure}[b]{0.32\textwidth}
         \centering
         \includegraphics[width=\textwidth]{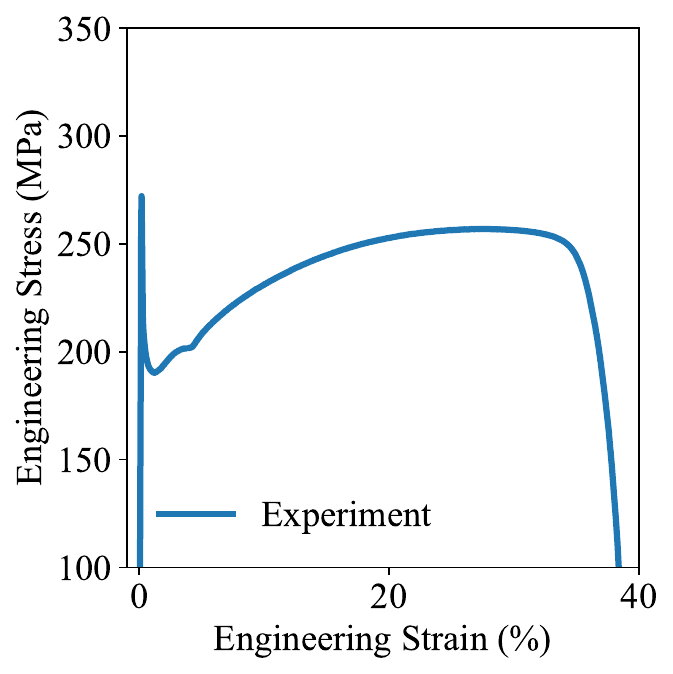}
         \caption{}
         \label{fig:mat_model_eng_stress}
     \end{subfigure}
     \hfill
     \centering
     \begin{subfigure}[b]{0.32\textwidth}
         \centering
         \includegraphics[width=\textwidth]{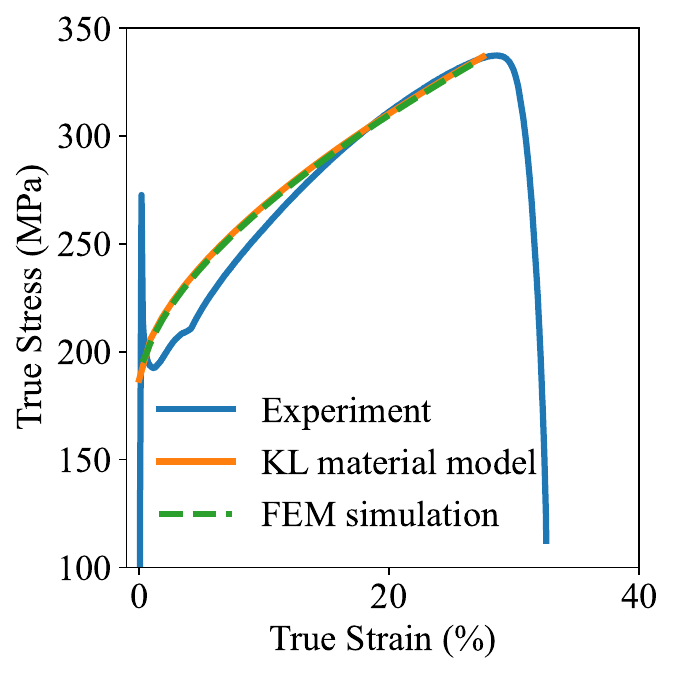}
         \caption{}
         \label{fig:mat_model_true_stress}
     \end{subfigure}
     \hfill
     \caption{Tensile test: (a) Displacement (top) and equivalent stress (bottom) distribution. (b) Engineering stress-strain curve and (c) True stress-strain curve.}
        \label{fig:mat_model_results}
\end{figure}

The resulting displacement field and equivalent Von Mises stresses obtained from the simulation of the tensile test are depicted in Figure \ref{fig:mat_model_tensile}. Here it can be seen that the stress distribution is homogeneous along the flat region of the dog-bone sample, giving a maximum value of 321 MPa for the imposed displacement of 5.7 mm. Then, the engineering stress-strain curve registered during the experimental test campaign is shown in Figure \ref{fig:mat_model_eng_stress}. The obtained curve is characterized by the presence of a pronounced upper yield strength, caused by the strain-rate sensitivity typical in the body-centered cubic (BCC) metals \cite{Prime2022}, followed by a L{\"u}ders-like region \cite {Colas2014} before transitioning to the strain hardening part, proceeded by necking and failure zones, respectively. Next, the experimental curve is transformed to the true-stress strain curve (see Figure \ref{fig:mat_model_true_stress}) where it is confirmed the good performance of the selected material model in the large strain region, that corresponds to our zone of interest during manufacturing. The comparison is obtained by superimposing the curves obtained by the analytical expression of the simplified KL material model and the FEM simulation, respectively. As a remark, the maximum true strain registered during the tensile test was 32.56 $\% $. This value provides an upper limit to the manufacturing process, in line with the estimated elongation of 31.75 $\%$ obtained in Section \ref{sec:Compression_bending}.

\subsection{Numerical model}
\label{ssec:Numerical_model}
In order to keep consistency with the experimental results, the same notation is used for the elbow cross section (see Figure \ref{fig:micrograph_test}), referred as LD and TD for the longitudinal and transverse directions, respectively. The results in terms of stresses and strains are referred to both regions. 

\begin{figure}[!ht]
     \includegraphics[width=\textwidth]{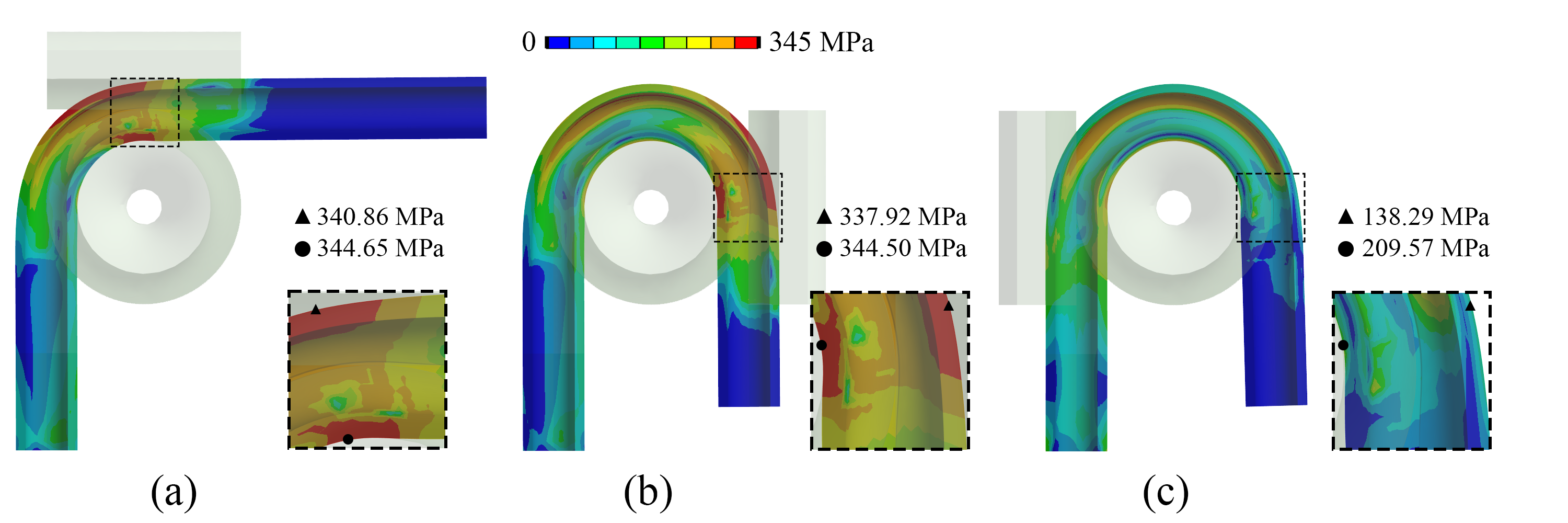}
     \caption{Elbow LD cross section: distribution of Equivalent stresses (Von Mises) during plastic bending (a) loading at 90$^\circ$ (b) loading at 180$^\circ$ and (c) residual stresses after unloading.}
        \label{fig:results_eqStress}
\end{figure}
\begin{figure}[!ht]
     \includegraphics[width=\textwidth]{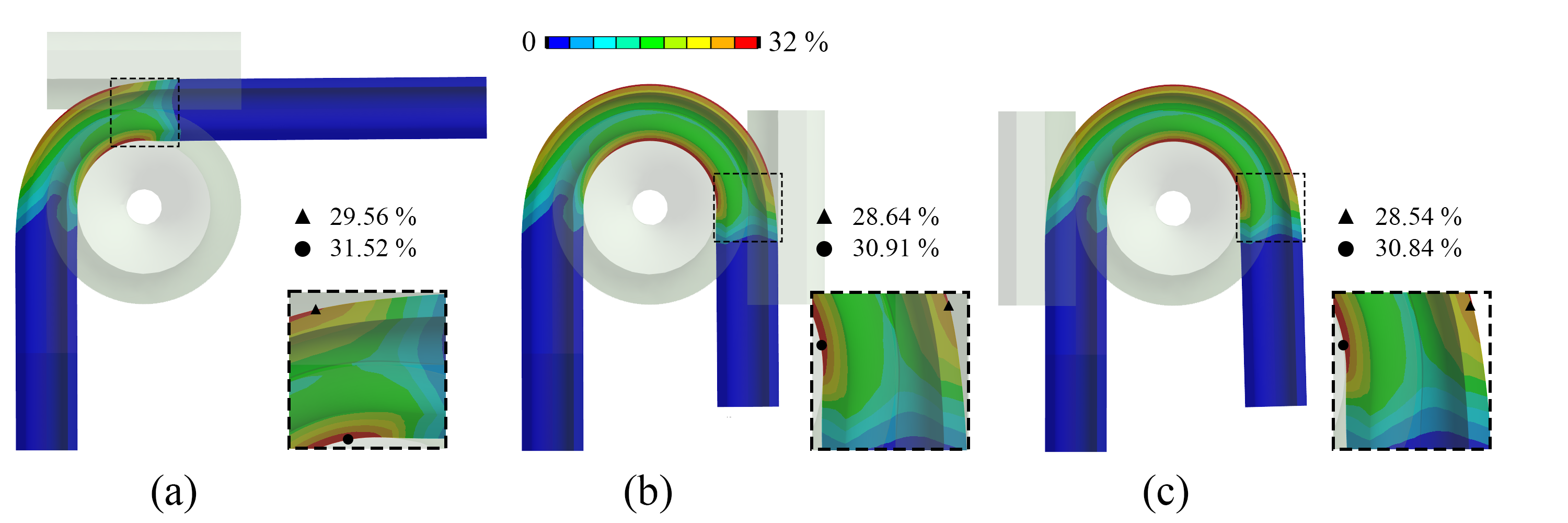}
     \caption{Elbow LD cross section: distribution of Equivalent total strains (Von Mises) during plastic bending (a) loading at 90$^\circ$ (b) loading at 180$^\circ$ and (c) residual strains after unloading.}
        \label{fig:results_eqStrain}
\end{figure}
\begin{figure}[!ht]
    \centering
         \includegraphics[width=\textwidth]{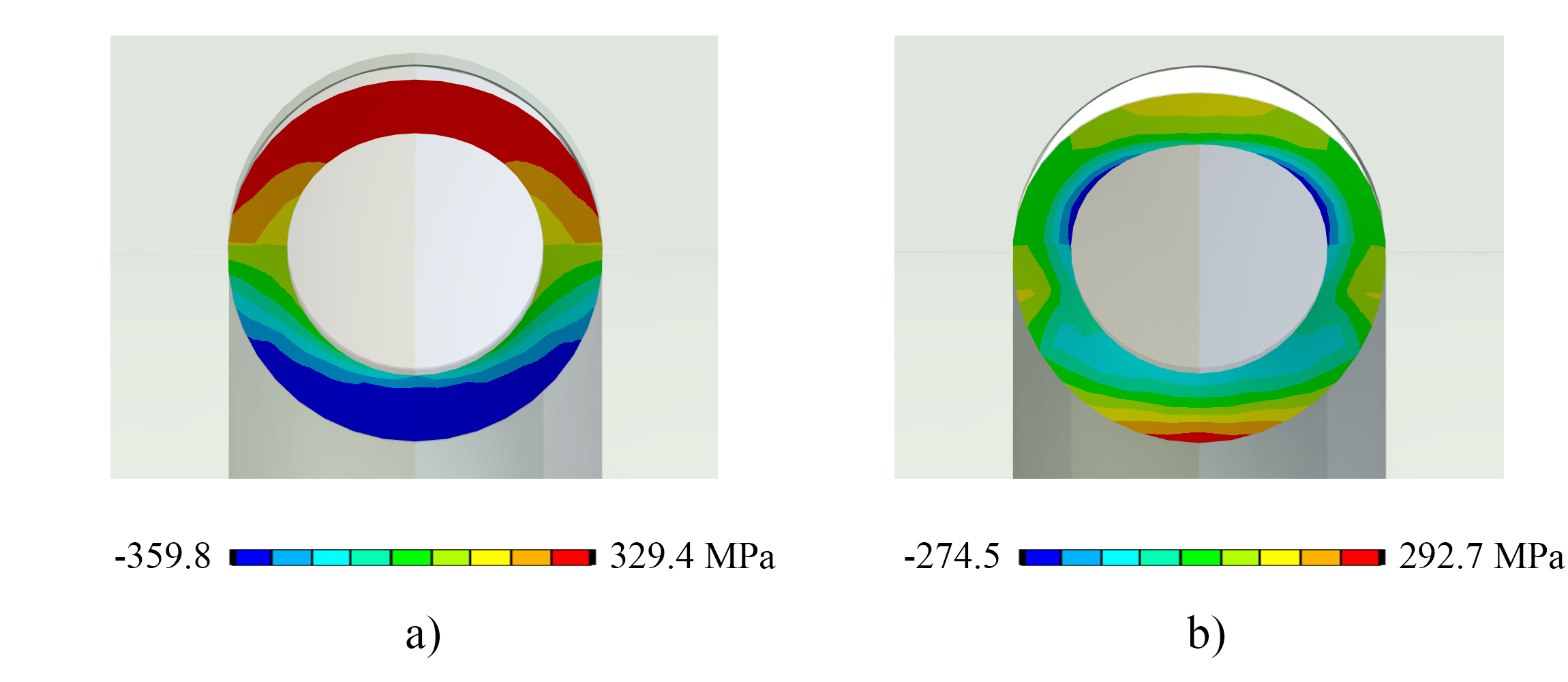}
     \caption{Elbow TD cross section: Longitudinal stresses $\sigma_z$ (a) during loading  and (b) resulting residual stresses after unloading.}
        \label{fig:FEM_sigma}
\end{figure}

The change of geometry as a consequence of the manufacturing process of the elbow leaves important residual stresses along the tantalum tube, which are inherent to the operation of plastic bending and are maximized when using the minimum allowable bending radius. Figure \ref{fig:results_eqStress} shows three snapshots of the process with the movable forming tool at 90$^\circ$, 180$^\circ$ and after unloading. Here, it can be seen how the material is subjected to stress hardening along the elbow with a maximum equivalent stress $\sigma_{eq}$ of 344.65 MPa located at the outer layer of the surface in contact with the fixed die at 90$^\circ$, marked with a dot in Figure \ref{fig:results_eqStress}a, while its counterpart (outer layer in contact with the moving die) presents a slightly lower value during loading and it is marked with a triangle. Note how the stress distribution is similar when completing the loading step (Figure \ref{fig:results_eqStress}b) and it is rotated the remaining 90$^\circ$. This is caused as each slice of the cross section experiences a similar state of equivalent stresses. Then, the process is completed after unloading. Here it can be seen how the the residual stresses are developed (see Figure \ref{fig:results_eqStress}c) but now, there is an important difference between both external layers, giving a higher residual stress again in the concave surface with a value of 209.57 MPa, while the convex counterpart presents 138.29 MPa, giving a difference of 34\%. To explain better this mismatch, Figure \ref{fig:results_eqStrain} shows the equivalent total strains $\varepsilon_{eq}$. Here it can be seen how homogeneous the strain distribution is with the maximum values located at both external layers, colored in red\footnote{The maximum equivalent strain value of 31.52\% experienced during loading (see Figure \ref{fig:results_eqStrain}a) goes in line with the values reported in the material model (see subsection \ref{ssec:Material_model})}. However, as the total strains can be decomposed in an elastic and plastic contribution, $\varepsilon_{eq}=\varepsilon_{eq}^e+\varepsilon_{eq}^p$, using the values depicted in Figure \ref{fig:results_eqStrain}b-c, it is found that the elastic strain for the convex surface is higher than the concave region, with values of 0.1 and 0.07\%, respectively. 

\begin{figure}[!htb]
    \centering
     \begin{subfigure}[b]{0.63\textwidth}
         \centering
         \includegraphics[width=\textwidth]{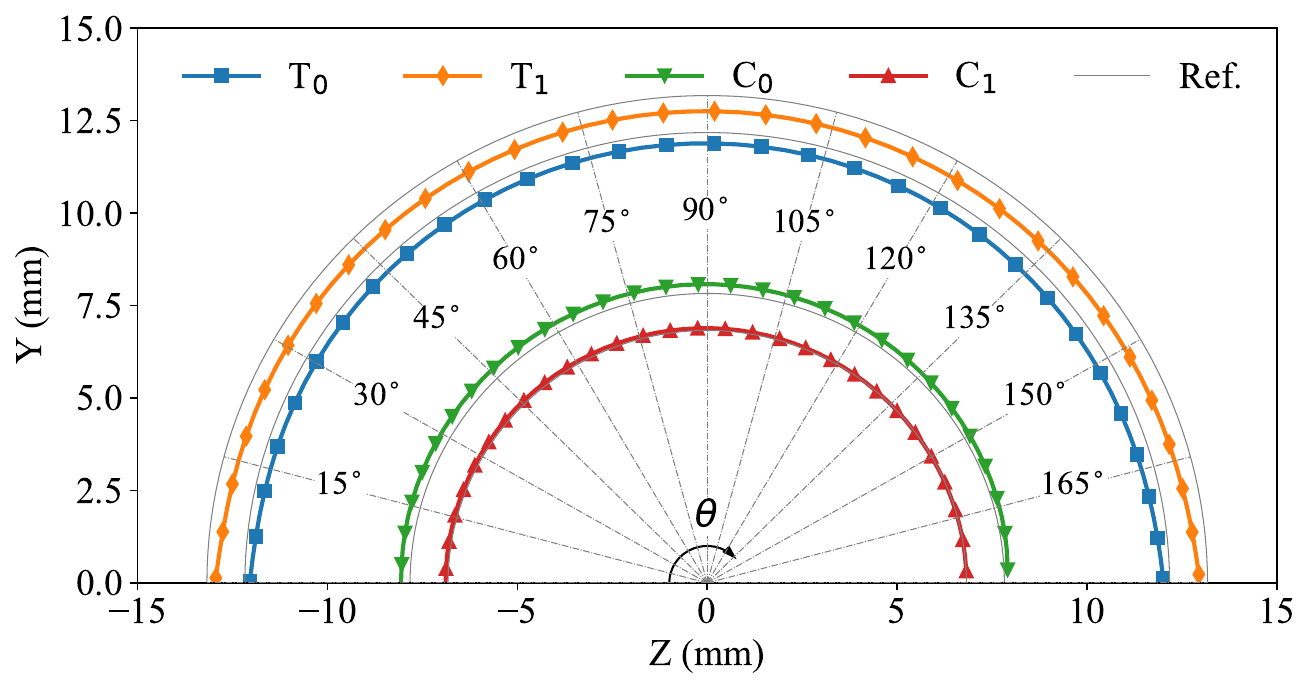}
         \caption{}
         \label{fig:Elbow_LD_FEM_distortion_a}
     \end{subfigure}
     \begin{subfigure}[b]{0.36\textwidth}
         \centering
         \includegraphics[width=\textwidth]{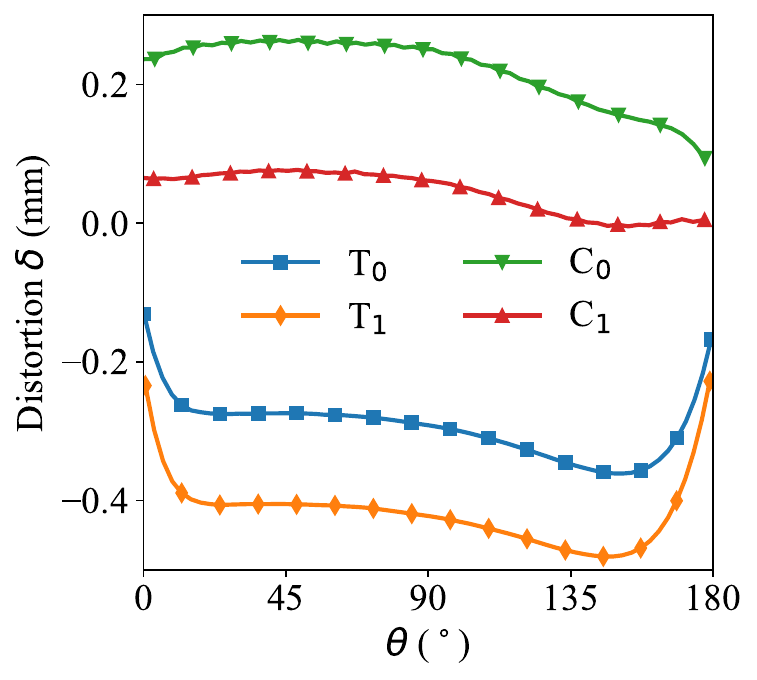}
         \caption{}
         \label{fig:Elbow_LD_FEM_distortion_b}
     \end{subfigure}
     \caption{Elbow LD cross section: (a) comparison between the theoretical geometry and the resulting radii and (b) distortion distribution $\delta$ as a function of the elbow angle $\theta$.}
        \label{fig:Elbow_LD_FEM_distortion}
\end{figure}

While the use of the equivalent stresses discussed above is useful to explain the plastic phenomenon once the yield stress is overcome, which for pure tantalum is $\sigma_y$ = 187.02 MPa, the positive scalar nature of $\sigma_{eq}$ hides the involved tension and compression stresses. To turnaround this detail, Figure \ref{fig:FEM_sigma} shows the longitudinal stresses $\sigma_z$ at the elbow TD cross section, as it corresponds to the main stress contribution, based on the beam theory. Note that this specific location coincides to the cross section of the elbow at 90$^\circ$. During loading (Figure \ref{fig:FEM_sigma}a), it can be seen that the stress distribution between the external regions is not even, where the portion working under compression is subjected to higher stresses (in absolute terms) with respect to its counterpart in tension with values of -359.8 and 329.4 MPa, respectively. Then, after completing the loading and half way the unloading cycle (Figure \ref{fig:FEM_sigma}b), the reversion in stress pattern takes places, leaving the lower external surface working under tension while a portion of the upper internal surface is under compression, with values of 292.7 and -274.5 MPa, respectively.

The difference in stresses levels experienced by the lower and upper regions is caused by the increment in thicknesses of the lower region working under compression and a reduction in the portion subjected to tension, which is the precursor of the cross section distortion refereed as ovalization. As a result, the neutral axis shift downwards, in direction to the portion under compression. The distortion respect to the theoretical radius evolves along the process. In terms of ovalization, although the process is not finished, in Figure \ref{fig:FEM_sigma}a the gap between the theoretical shape and the resulting distorted cross section is observed already. After unloading, as ovalization was completed, the resulting gap is bigger, as shown in Figure \ref{fig:FEM_sigma}b. To report the distortion along the elbow, the LD cross section is used. To this end, the theoretical and final geometry are plotted in Figure \ref{fig:Elbow_LD_FEM_distortion_a}. To properly name each of the four radii, the following nomenclature is used: $T$ and $C$ comes for tension and compression and the subscripts 0 and 1 correspond to the internal and external faces, respectively. For example, $C_1$ refers to the external radius of the region working under compression. The four resulting curves differ from the theoretical geometry. As expected, the thickening of the compression side (see $C_1$) and the narrowing of the tension side (see $T_0$ and $T_1$) are obtained. Once the curves are know in the final configuration, the radial distortion $\delta$ for each geometry is calculated as:

\begin{equation}
    \delta(\theta) = R_{ref} - R_{sim}(\theta)
\end{equation}

where $R_{ref}$ and $R_{sim}$ correspond to the reference radius (theoretical) and the radius obtained in the simulation after unloading, respectively. The resulting distortion as a function of the elbow angle $\theta$ is shown in Figure \ref{fig:Elbow_LD_FEM_distortion_b}. Here, it can be seen that distortion is present in all four radii where the main difference occurs in T$_1$ with a value of -0.481 mm at 148.3$^\circ$.

At a simulation level, for the given tool rotation of 180$^\circ$, the obtained final elbow angle is 178.458$^\circ$. Therefore, springback compensation is needed.

\subsection{Experimental campaign}
\label{ssec:Experimental_campaign}

\subsubsection*{Metrology}
\begin{figure}[!htb]
    \centering
         \includegraphics[width=\textwidth]{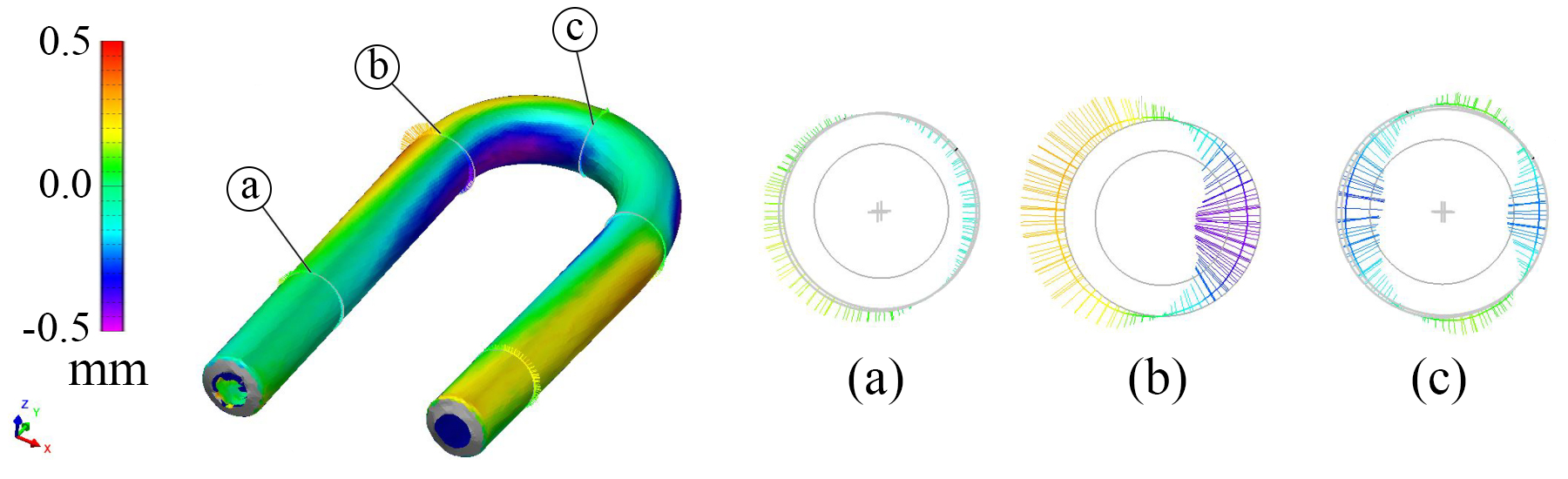}
         \caption{Metrology measurements: (left) distortion field on the sample and (right) cross section at (a) below, (b) beginning and (c) half elbow.}
        \label{fig:metrology_results}
\end{figure}

Figure \ref{fig:metrology_results} shows the deviation from the obtained geometry respect to the nominal shape. The variation was bounded in a range of $\pm$ 0.5 mm. The measured outer diameter was 6.395 mm. Therefore, the circularity tolerance in the nominal portion of the tube was +45 $\mu$m (see cross section (a)), showing that the tube was slightly bigger. This value falls within the tolerance of 0.102 mm prescribed by the ASTM B521 R05200 manufacturing standard \cite{ASTMB521}. Then, ovalization was registered from the beginning of the elbow (see cross section (b)) where a strong distortion with a w-shape was observed in the compression side, counterbalanced by a more homogeneous and progressive c-shape warped half circle on the tension side. The peak values for each portion were -0.41 and +0.29 mm, respectively. Finally, the flatness caused by the reduction/increment of thickness at the tension/compression side and slight lateral expansion was confirmed in cross section (c) by registering an inscribed circle of diameter 6.194 mm. The same section was measured a second time with a caliper, and the  maximum/minimum distance was 6.43/5.87 mm, respectively. The information provided in this test corresponds only to the external surface of the tube, i.e no thickness measurements were performed.

\subsubsection*{Cross-section and microstructure samples}
\begin{figure}[!hb]
    \centering
     \begin{subfigure}[b]{0.31\textwidth}
         \centering
         \includegraphics[width=\textwidth]{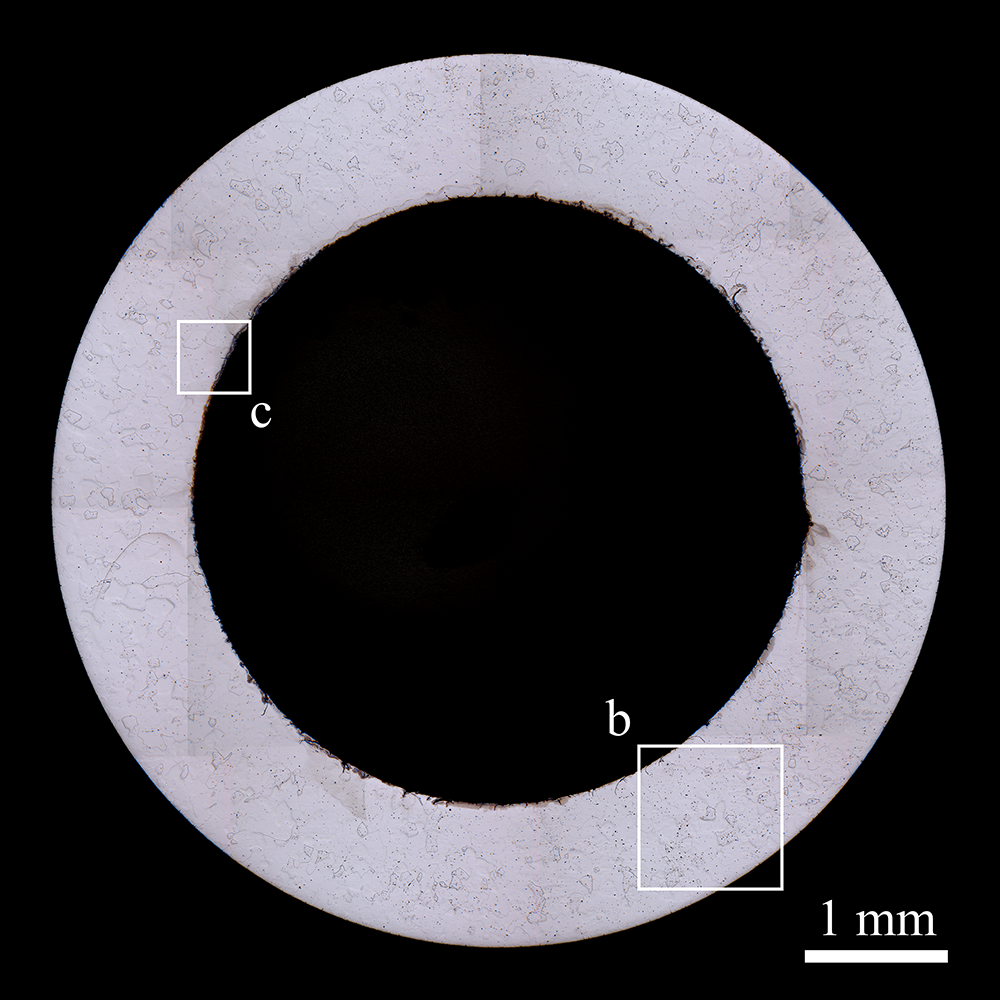}
         \caption{}
         \label{fig:Tube_TD_micro}
     \end{subfigure}\quad
     \begin{subfigure}[b]{0.31\textwidth}
         \centering
         \includegraphics[width=\textwidth]{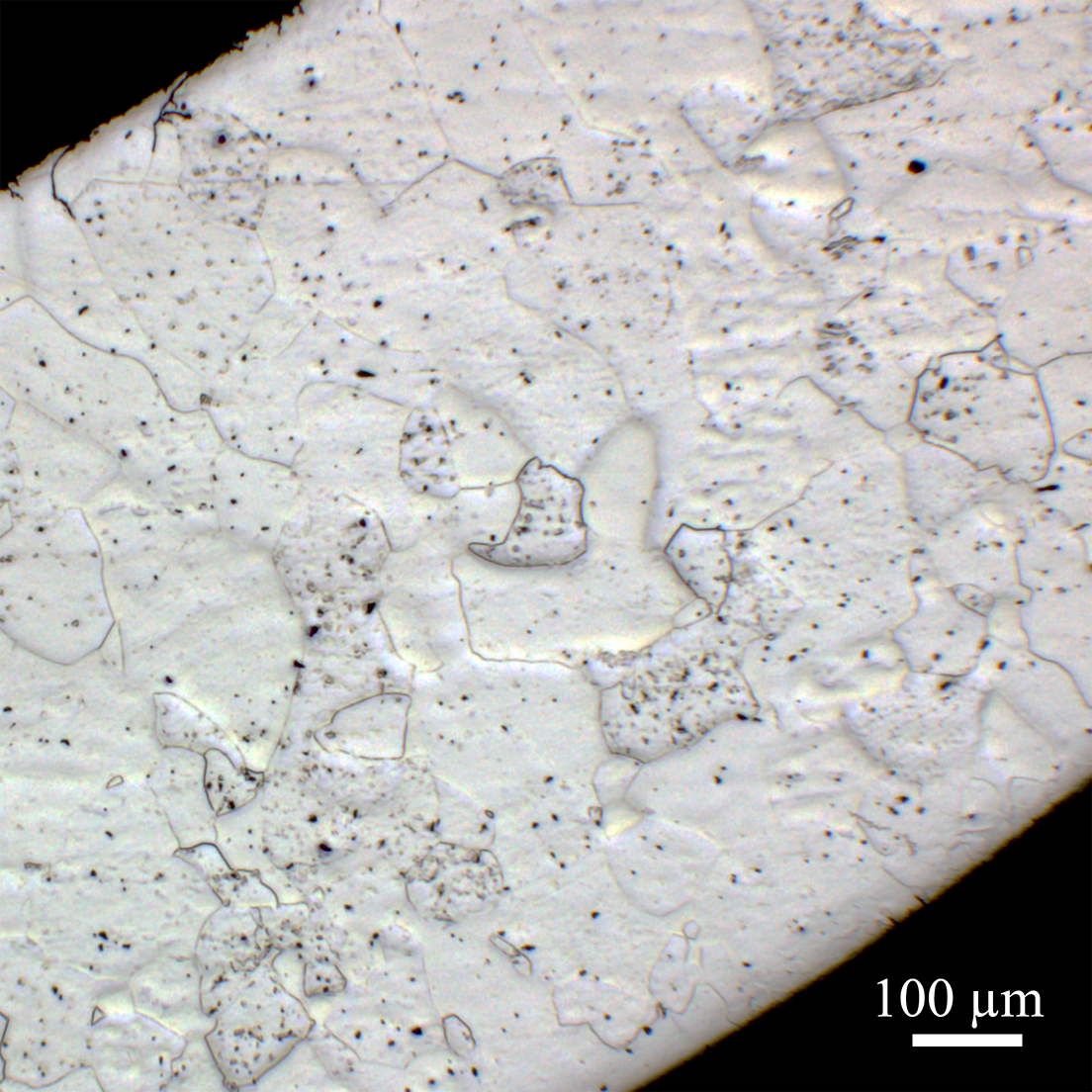}
         \caption{}
         \label{fig:Tube_TD_structure}
     \end{subfigure}\quad
     \begin{subfigure}[b]{0.31\textwidth}
         \centering
         \includegraphics[width=\textwidth]{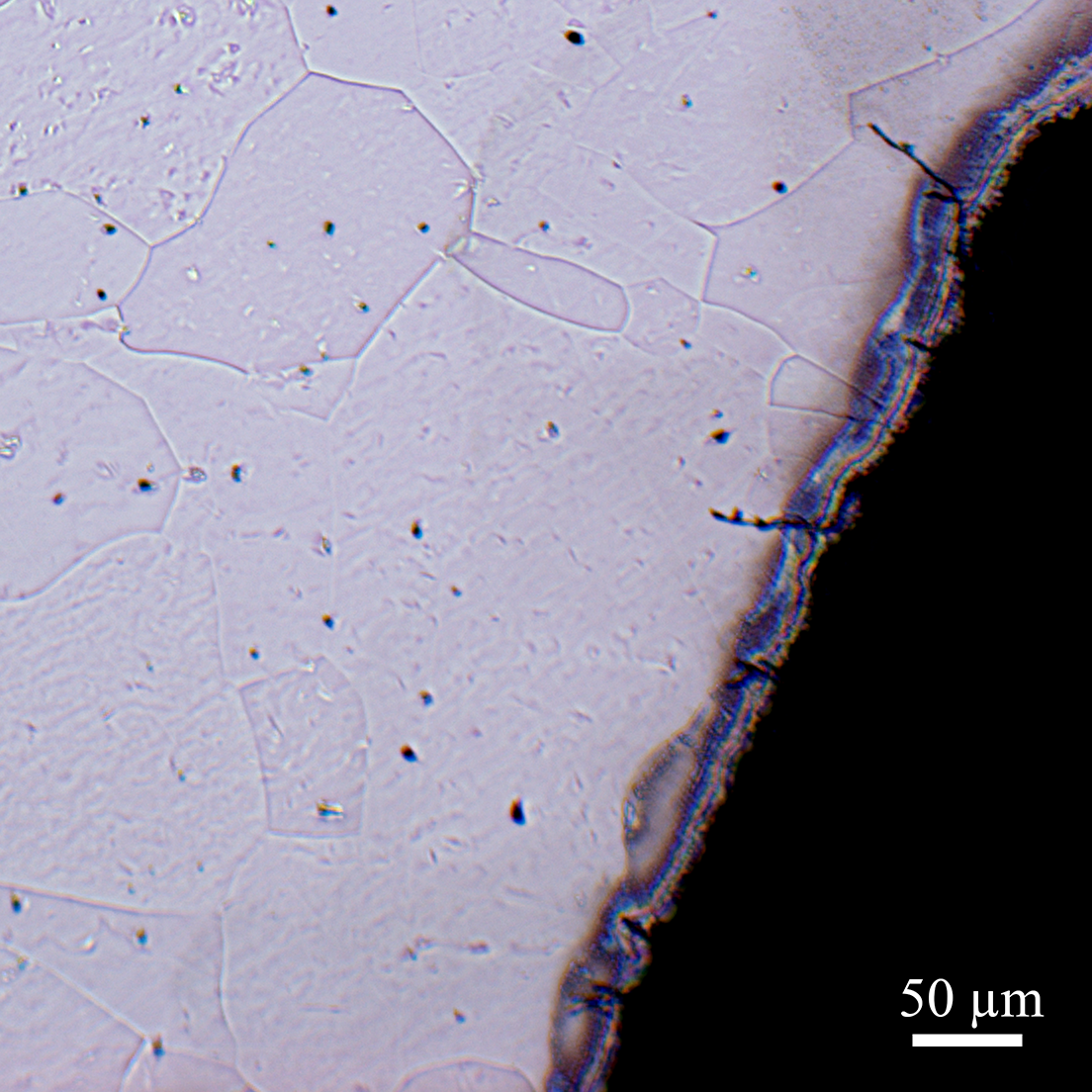}
         \caption{}
         \label{fig:Tube_TD_oxide}
     \end{subfigure}
     \caption{Tube TD cross section: (a) nominal geometry, (b) grain size micrograph and (c) presence of an oxide layer and micro-cracks along the internal wall.}
        \label{fig:Tube_TD}
\end{figure}

Figure \ref{fig:Tube_TD_micro} shows the nominal cross-section in the transverse direction of the tantalum tube, with an average wall thickness of 1.022 $\pm$ 0.173 mm.
By zooming the picture, the microstructure before cold-working is depicted in Figure \ref{fig:Tube_TD_structure}. Here it can be seen the polycristalline nature of the sample with an homogeneous grain distribution, and an  equivalent grain diameter of 79 $\mu$m. However, the internal wall of the tube presents a fine layer of oxide of 25 $\mu$m thickness and some microcracks with an average and maximum size of 9.76 and 112.17 $\mu$m, respectively, as shown in Figure \ref{fig:Tube_TD_oxide}. In addition, the presence of some micro-porosities along the sample is observed.

\begin{figure}[!ht]
    \centering
     \begin{subfigure}[b]{0.31\textwidth}
         \centering
         \includegraphics[width=\textwidth]{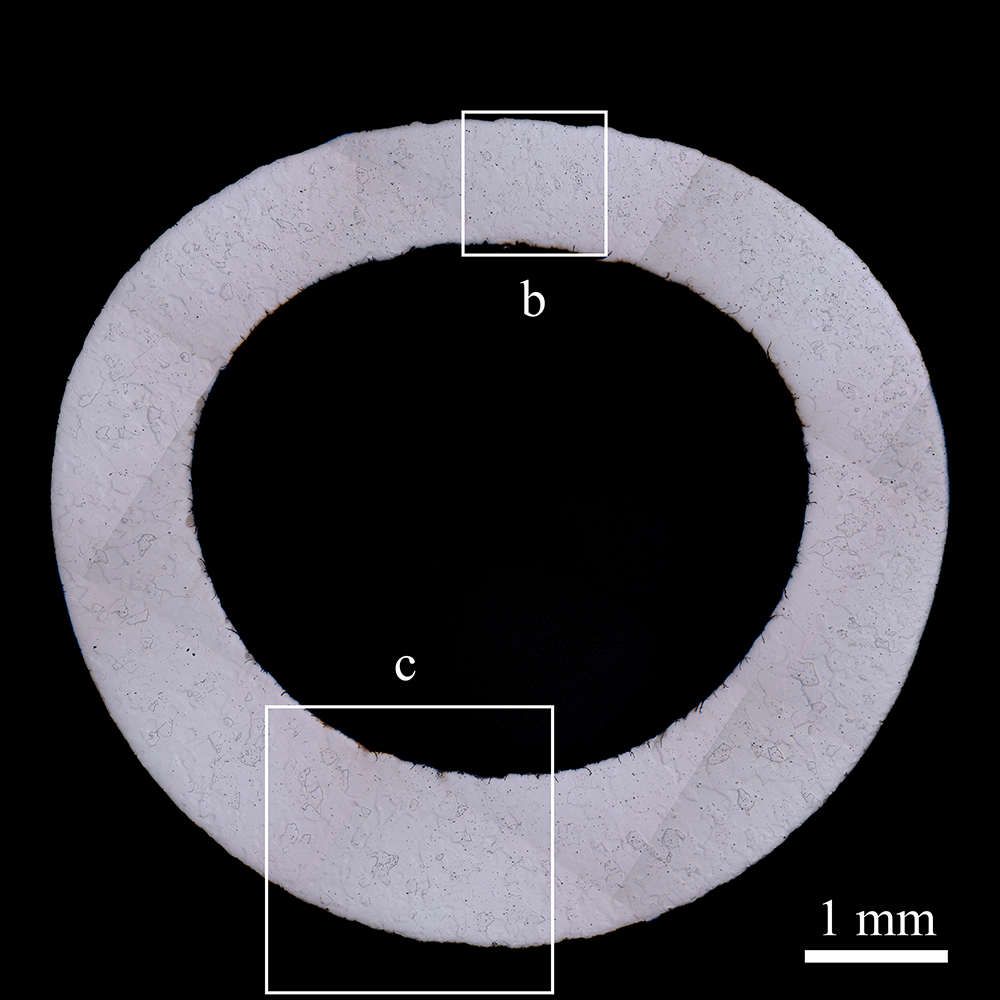}
         \caption{}
         \label{fig:Elbow_TD_micro}
     \end{subfigure}\quad
     \begin{subfigure}[b]{0.31\textwidth}
         \centering
         \includegraphics[width=\textwidth]{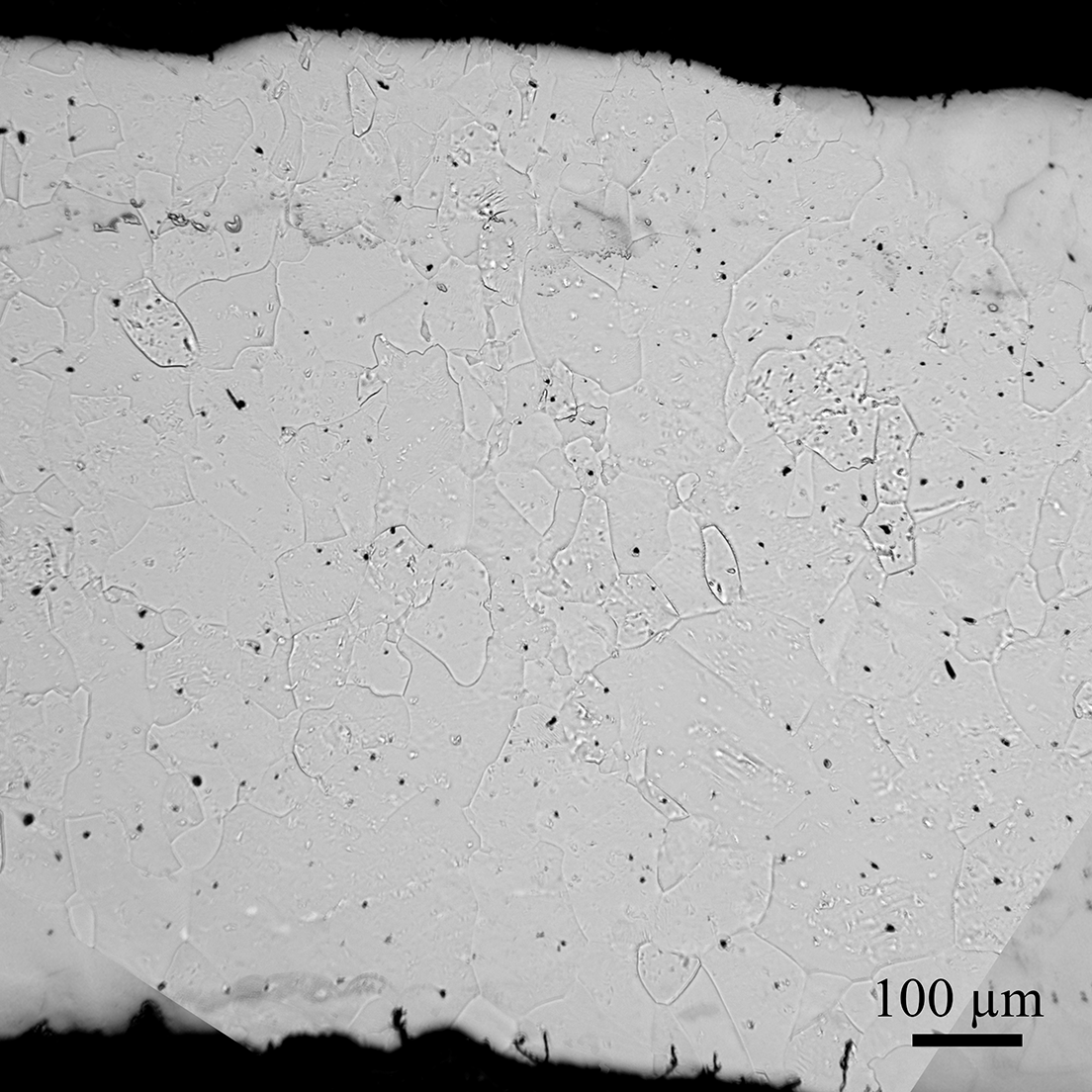}
         \caption{}
         \label{fig:Elbow_TD_ten}
     \end{subfigure}\quad
     \begin{subfigure}[b]{0.31\textwidth}
         \centering
         \includegraphics[width=\textwidth]{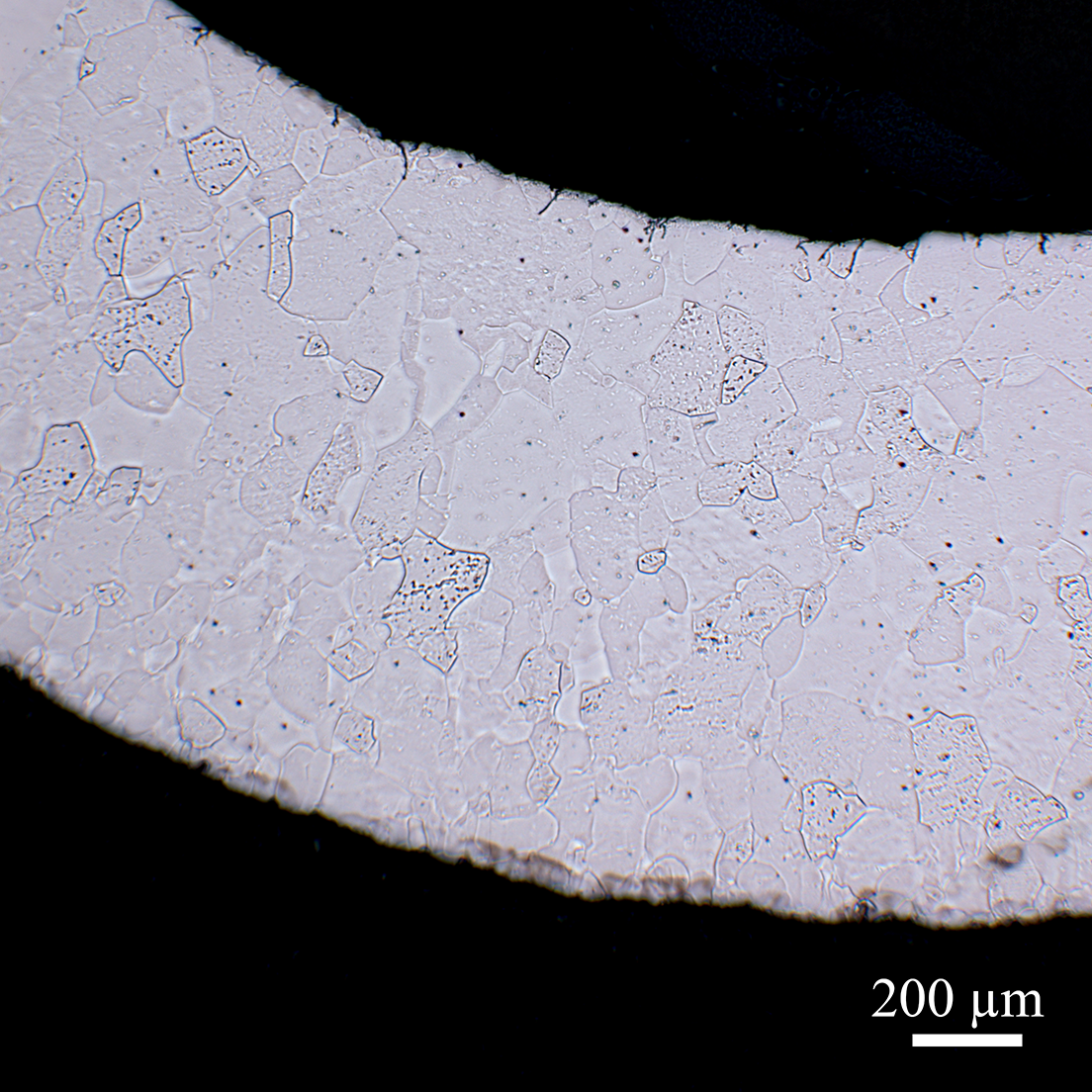}
         \caption{}
         \label{fig:Elbow_TD_comp}
     \end{subfigure}
     \caption{Elbow TD cross section: (a) resulting distorted geometry, (b) tension  and (c) compression details for grain size comparison.}
        \label{fig:Elbow_TD}
\end{figure}

\begin{figure}[!ht]
    \centering
     \begin{subfigure}[b]{0.31\textwidth}
         \centering
         \includegraphics[width=\textwidth]{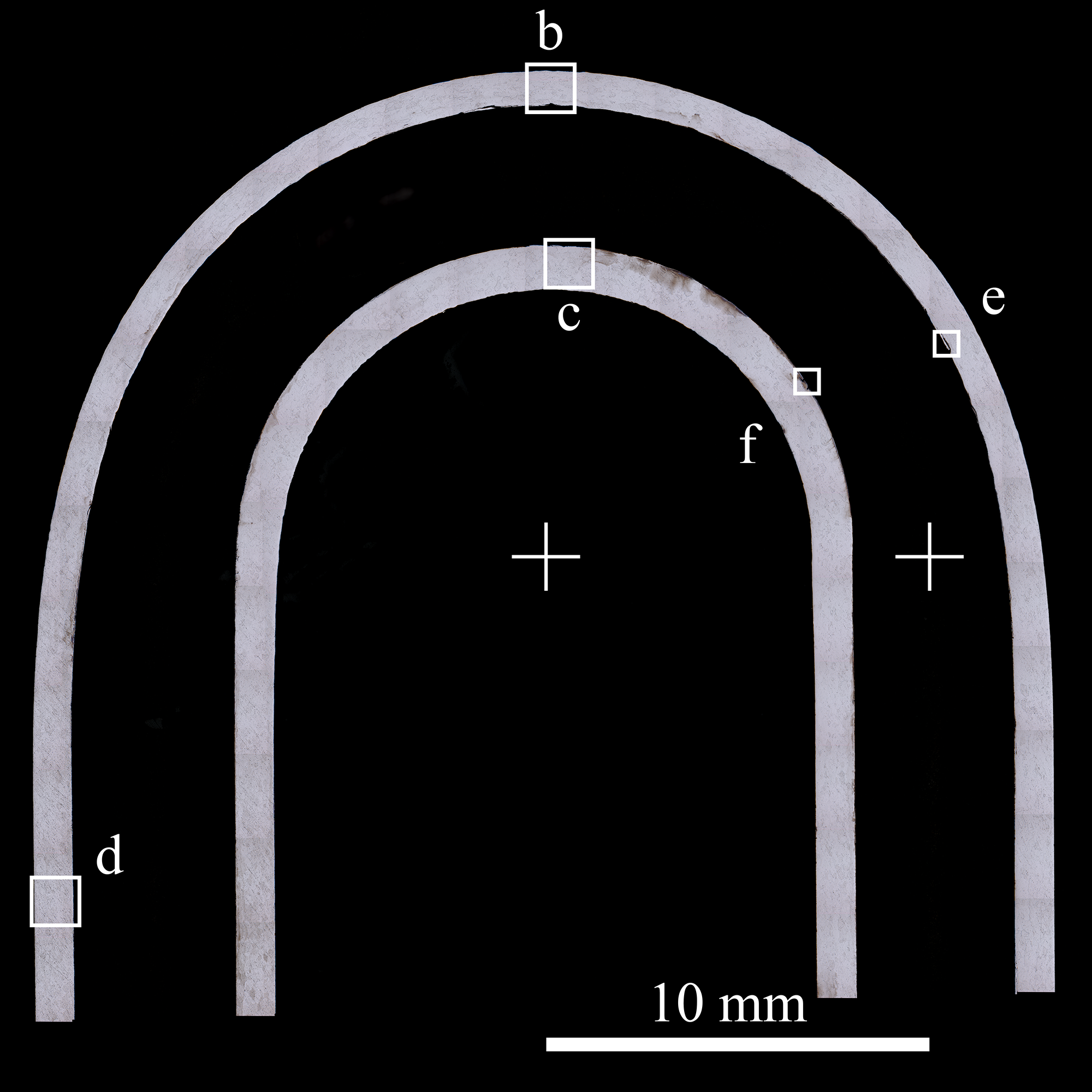}
         \caption{}
         \label{fig:Elbow_LD_a}
     \end{subfigure}\quad
     \begin{subfigure}[b]{0.31\textwidth}
         \centering
         \includegraphics[width=\textwidth]{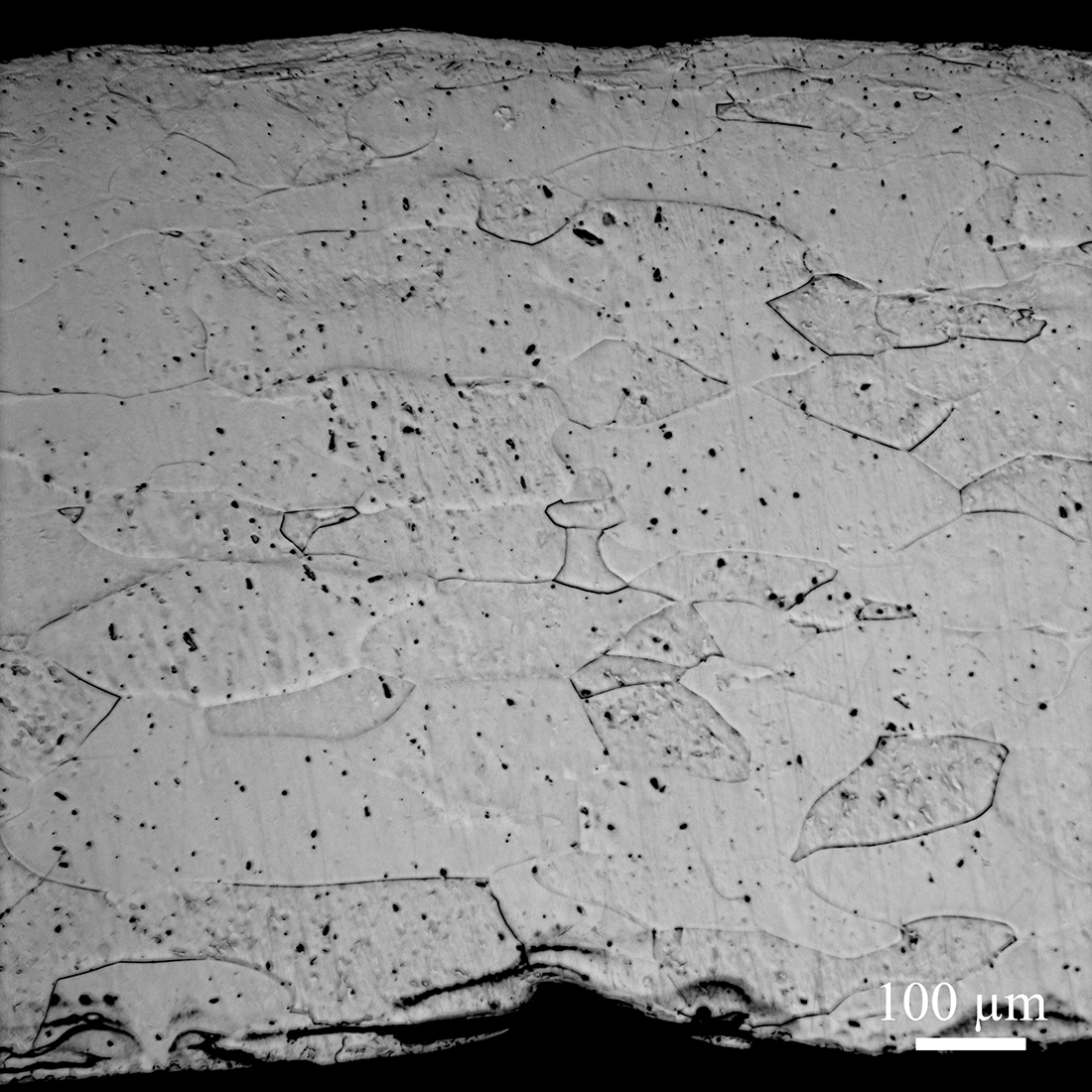}
         \caption{}
         \label{fig:Elbow_LD_ten}
     \end{subfigure}\quad
     \begin{subfigure}[b]{0.31\textwidth}
         \centering
         \includegraphics[width=\textwidth]{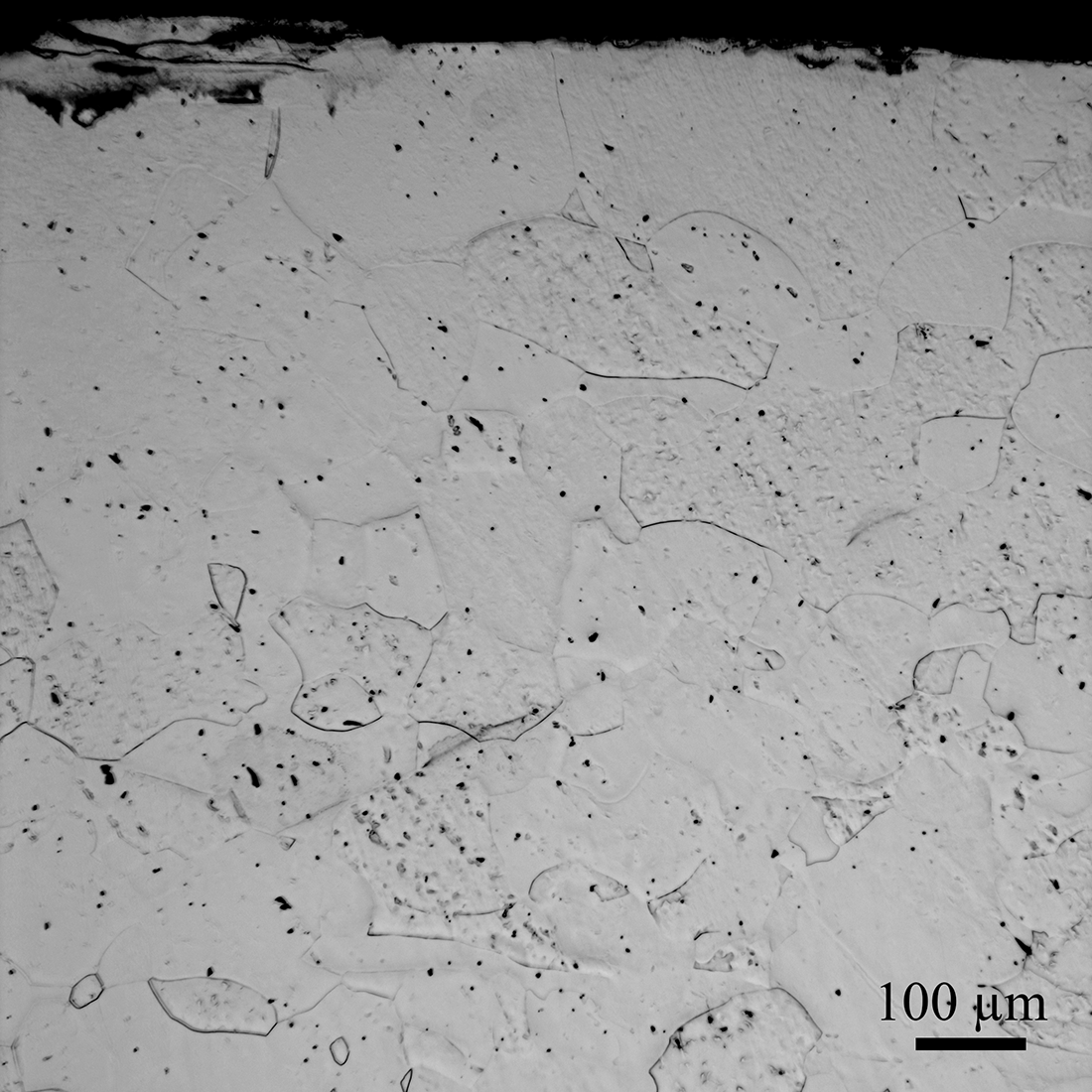}
         \caption{}
         \label{fig:Elbow_LD_comp}
     \end{subfigure}\quad
     \begin{subfigure}[b]{0.31\textwidth}
         \centering
         \includegraphics[width=\textwidth]{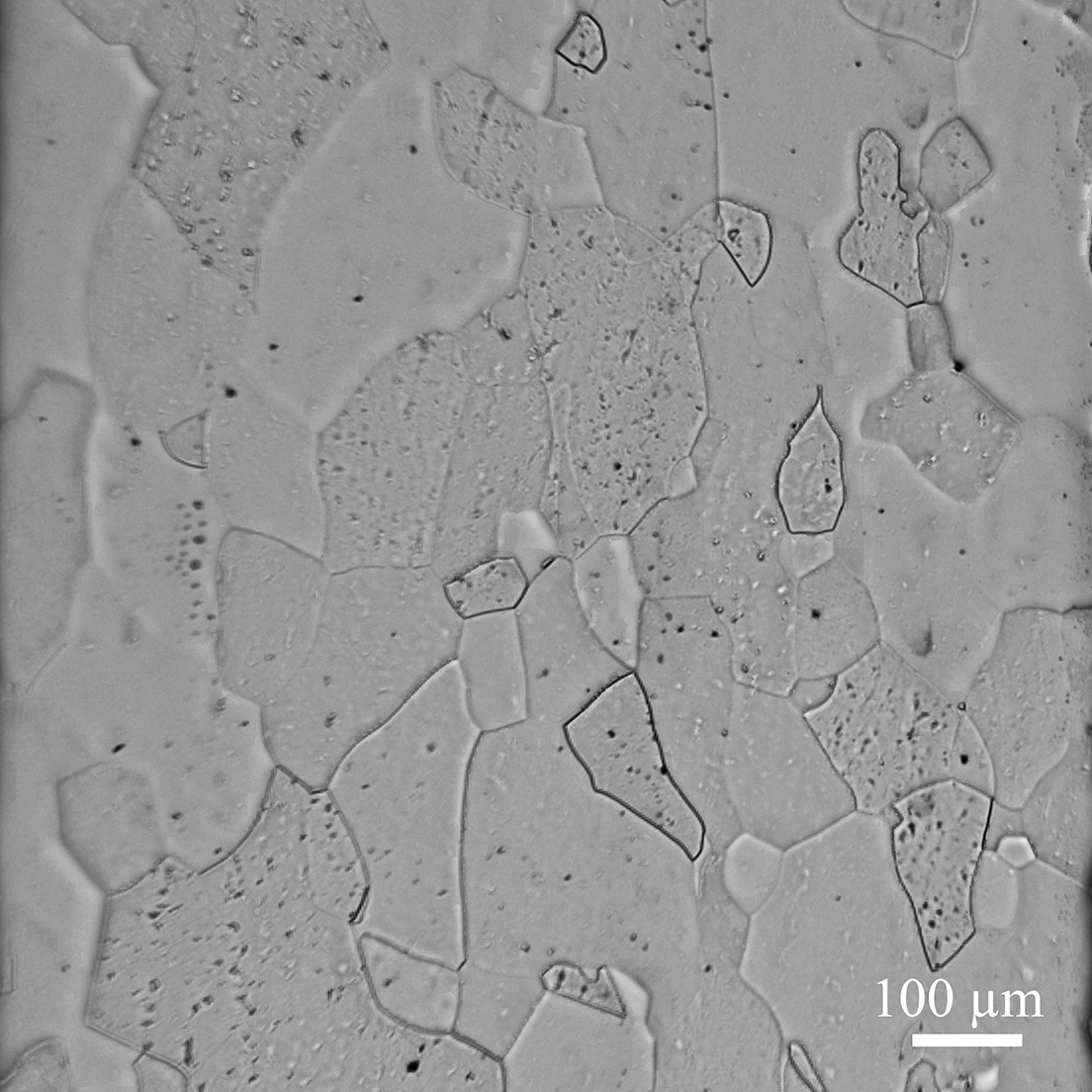}
         \caption{}
         \label{fig:Tube_LD}
     \end{subfigure}\quad
       \begin{subfigure}[b]{0.31\textwidth}
         \centering
         \includegraphics[width=\textwidth]{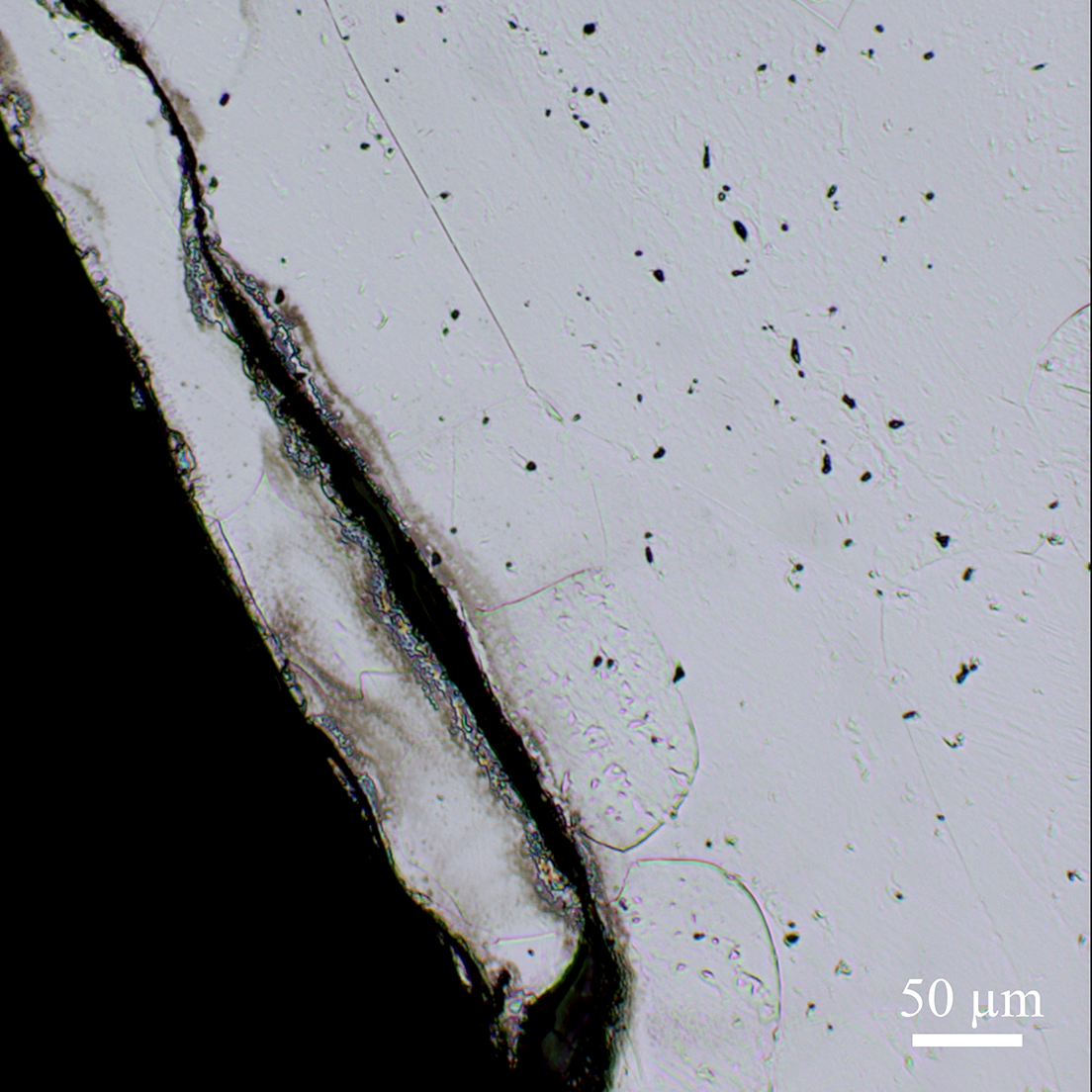}
         \caption{}
         \label{fig:Elbow_LD_ox_ten}
     \end{subfigure}\quad
       \begin{subfigure}[b]{0.31\textwidth}
         \centering
         \includegraphics[width=\textwidth]{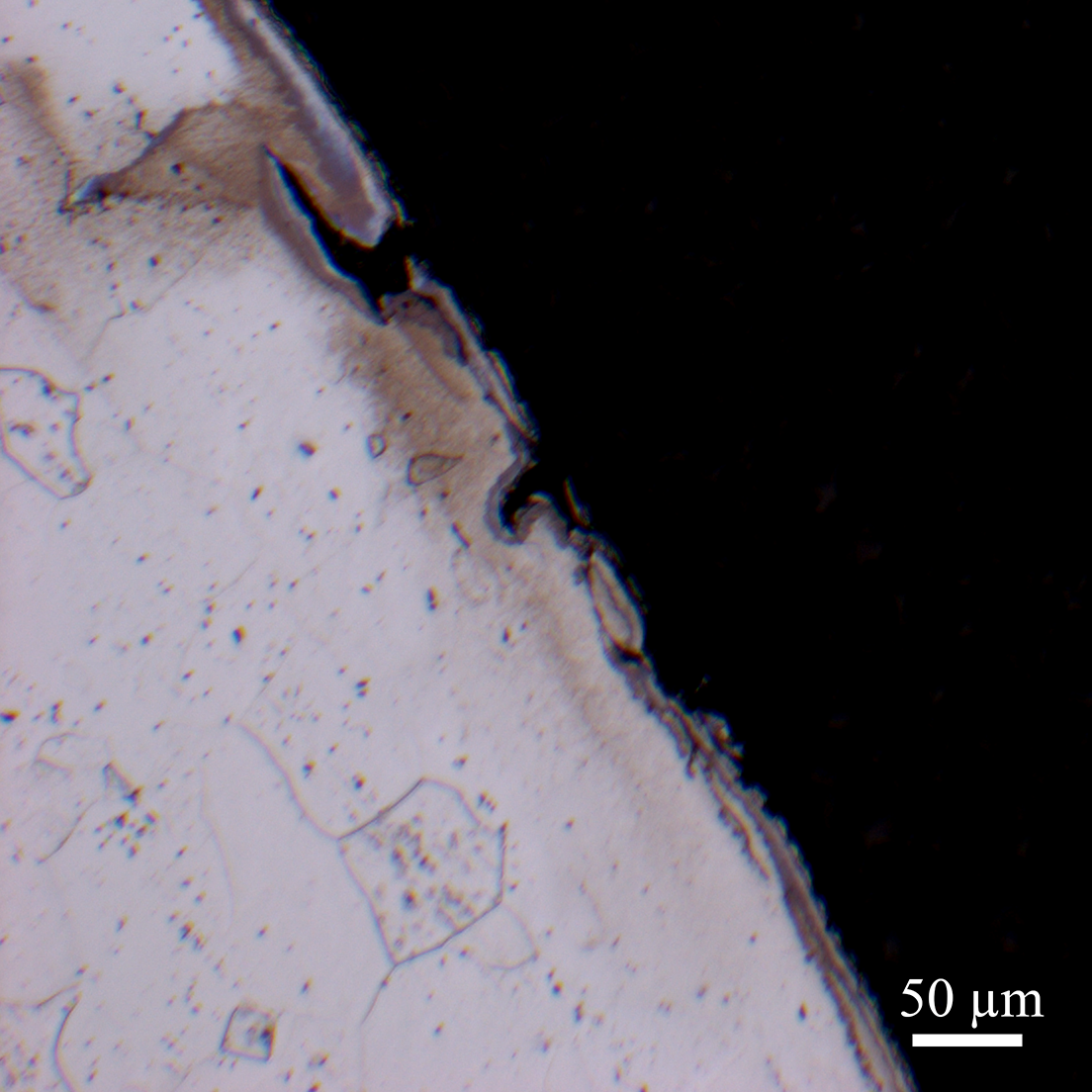}
         \caption{}
         \label{fig:Elbow_LD_ox_comp}
     \end{subfigure}
     \caption{Elbow LD cross section: (a) thickness evolution profile, (b) outer radius  and (c) inner radius details, (d) grain size micrograph in the pristine region, presence of oxide layers in the internal walls for the (e) outer radius and (f) inner radius, respectively.}
        \label{fig:Elbow_LD}
\end{figure}

On the other hand, Figure \ref{fig:Elbow_TD_micro} presents the resulting cross-section of the elbow, where the effects of ovalization are fully visible. The deformed geometry can be enclosed by a rectangle of dimensions 6.399 $\times$ 5.916 mm. The values correspond to the width and height, respectively and they go in line with the results obtained before by metrology. In addition, the wall thickness in the tension/compression side are 0.913/1.26 mm, respectively. Moreover, the cold-working effects can be seen in the form of an elongated grain along the radial direction, especially in the tension side, as depicted in Figure \ref{fig:Elbow_TD_ten}. Again, the presence of microcracks in the internal wall are confirmed, as shown in Figures \ref{fig:Elbow_TD_ten} and \ref{fig:Elbow_TD_comp}. 

Figure \ref{fig:Elbow_LD_a} presents the resulting longitudinal cross-section of the elbow, where the progressive reduction/increment of the wall thickness on the tensile/compressive walls is observed. In the micrography, five sections were augmented to observe different features of the (un)deformed tube. Starting with the grain morphology, while the outer radius, subjected to tensile stresses presents a more elongated grain geometry in the axial direction (see Figure \ref{fig:Elbow_LD_ten}), the inner radius has a more homogeneous grain size distribution with some variation along the radial direction, as depicted in Figure \ref{fig:Elbow_LD_comp}. However, both cases present small defects in the internal walls of the tube (see middle bottom and upper left parts of Figures \ref{fig:Elbow_LD_ten} and \ref{fig:Elbow_LD_comp}, respectively) that correspond to the initial defects of the tube described before. In addition, the effects of the contact between the tool an the tube are observed on top of Figure \ref{fig:Elbow_LD_ten}, where a wavy surface is obtained with a layered shape beneath the external face. On the other hand, the pristine poly-crystalline nature of the seamless tube in the longitudinal direction is shown in Figure \ref{fig:Tube_LD}. Here, it can be seen the flakes with an average length of 125 $\mu$m for the major axis. In addition, the presence of oxide layers in the internal walls of the tube is depicted in Figures \ref{fig:Elbow_LD_ox_ten}-\ref{fig:Elbow_LD_ox_comp}, located at the tension and compression sides, respectively. 

\subsubsection*{Hardness measurements}

\begin{figure}[!htb]
    \centering
     \begin{subfigure}[b]{0.32\textwidth}
         \centering
         \includegraphics[width=\textwidth]{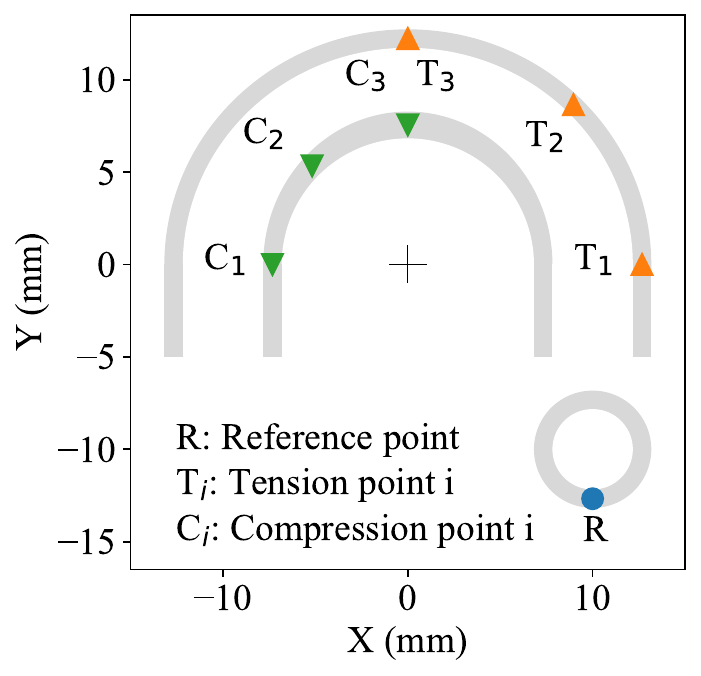}
         \caption{}
         \label{fig:HV_measurement_points}
     \end{subfigure}
     \centering
     \begin{subfigure}[b]{0.32\textwidth}
         \centering
         \includegraphics[width=\textwidth]{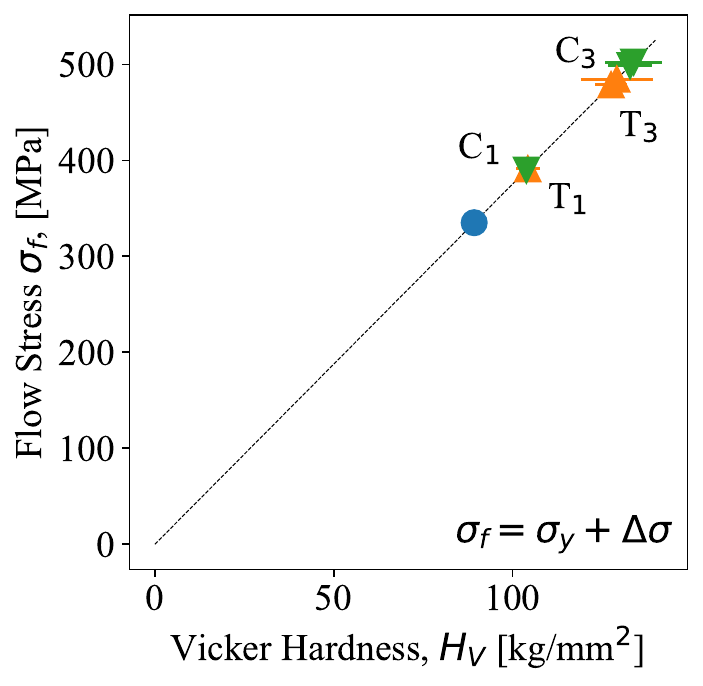}
         \caption{}
         \label{fig:sigma_f_HV}
     \end{subfigure}
     \begin{subfigure}[b]{0.32\textwidth}
         \centering
         \includegraphics[width=\textwidth]{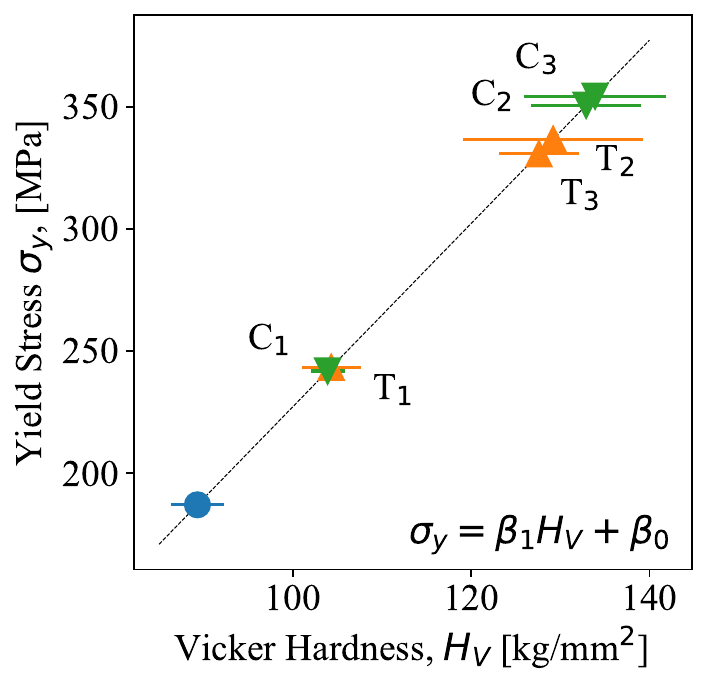}
         \caption{}
         \label{fig:Sigma_y_HV}
     \end{subfigure}
     \caption{Hardness measurements along the elbow: (a) Sampling points, (b) Flow Stress and (c) Yield stress estimations as a function of the Vicker Hardness.}
        \label{fig:Hardness}
\end{figure}

Figure \ref{fig:HV_measurement_points} shows the location of the six sampling points along the elbow LD cross section, together with the nominal hardness performed on the reference point (located on the tube TD cross section). The obtained numerical values are summarized on Table \ref{tab:hardness_raw_values}. From the measurements, it can be seen that hardness increased during the elbow deformation a 49.94\% due to cold-working  (from 89.3 $\pm$ 3.0 to 133.9 $\pm$ 8.0), being the maximum value registered on point $C_3$, that corresponds to the portion of the elbow at 90${^\circ}$ with the thickest wall. However, the upper part subjected to tension registered its maximum value at point $T_2$ (45${^\circ}$) instead to $T_3$.

From the hardness measurements, the flow stress $\sigma_f$ can be estimated following the procedure described by Tiryakio\v glu \cite{Tiryakioglu2015a}. Figure \ref{fig:sigma_f_HV} shows the results using the expression $\sigma_f =\frac{H_v}{0.927 C}$, where $H_v$ corresponds to the Vickers hardness (in MPa) and $C$ = 2.82 is the constrain factor. The obtained points were linearly fitted to $\sigma_f = \sigma_y+\Delta\sigma$, where $\sigma_y$ and $\Delta\sigma$ correspond to the yield stress and the increase in stress due to work hardening, respectively. Here, a perfect fit of the sampling points passing by origin was observed. 

Previous research shown that the yield stress can be expressed in the form  $\sigma_y = \beta_1H_v+\beta_0$ \cite{Tiryakioglu2015b}. Constant $\beta_1$ comes from the flow stress relation $\frac{1}{0.927C}$ $\approx$ 0.383 and the y-intercept constant $\beta_0$ was calculated using the yield stress obtained during the tensile test of 187.02 MPa, giving the relation $\sigma_y=0.383H_v-147.991$ for pure tantalum. The result is plotted in Figure \ref{fig:Sigma_y_HV}, where the maximum yield stress at $C_3$ is 354.337 MPa.

\begin{table}[!htb]
\centering
\caption{Vicker Hardness, H$_V$0.1 (kg/mm$^2$)}
\label{tab:hardness_raw_values}
\begin{tabular}{ cccc } 
 \hline
 \textbf{$i$} & \textbf{1} & \textbf{2} & \textbf{3}\\ 
 \hline
 R     &                 &  89.3 $\pm$  3.0  &\\
 T$_i$ & 104.3 $\pm$ 3.3 & 129.2 $\pm$ 10.1  & 127.6 $\pm$ 4.5\\
 C$_i$ & 103.9 $\pm$ 1.9 & 132.9 $\pm$  6.2  & 133.9 $\pm$ 8.0\\
 \hline
\end{tabular}
\end{table}

\section{Discussion}
\label{sec:Discussion}

The response of the material model presented in Section \ref{ssec:Material_model} is able to capture the behavior in the large strain regime, which is required for manufacturing purposes in the present study. However, it is worth mentioning that the material law is not capable to reproduce the upper yield registered in the stress-strain curve. In addition, during the first part of the strain-hardening regime, the material model provides an overestimation in term of stresses respect to the experimental curve. This limitation should be taken into account when using the simplified version of the KL model. As an alternative, Colas et al., \cite{Colas2014} developed a material model capable to capture the anomalous yield point in pure tantalum by using the KEMC (Kubin-Estrin-McCormick) phenomenological model. However, the model requires the use of 18 parameters.

\begin{figure}[!t]
    \centering
     \begin{subfigure}[b]{0.32\textwidth}
         \centering
         \includegraphics[width=\textwidth]{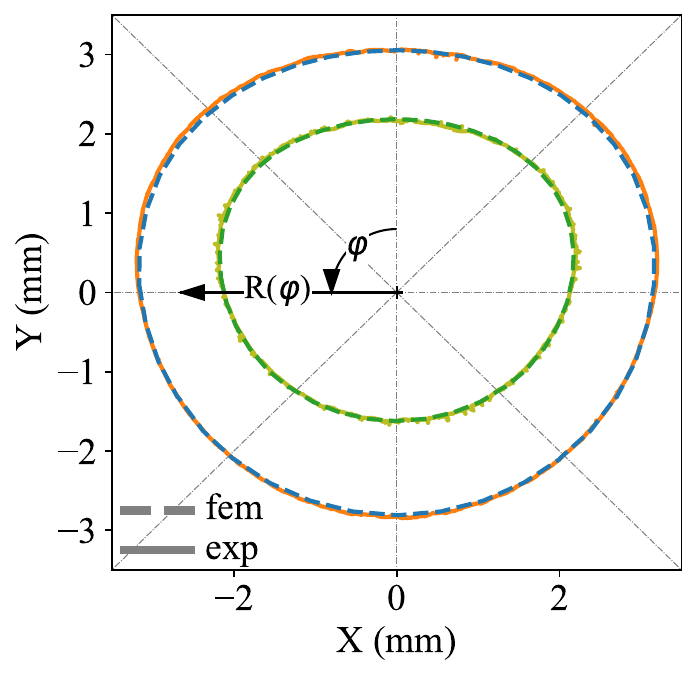}
         \caption{}
         \label{fig:elbow_TD_fem_exp}
     \end{subfigure}
     \centering
     \begin{subfigure}[b]{0.32\textwidth}
         \centering
         \includegraphics[width=\textwidth]{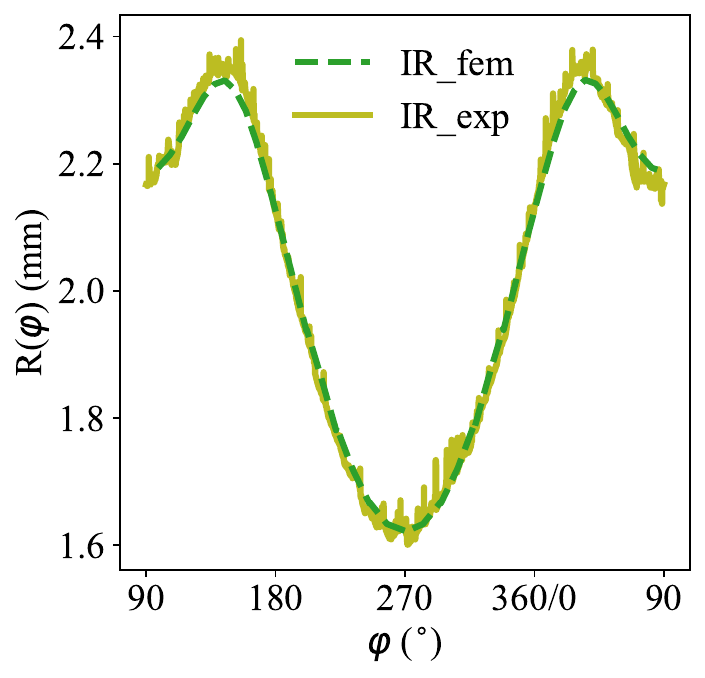}
         \caption{}
         \label{fig:elbow_TD_IR}
     \end{subfigure}
     \begin{subfigure}[b]{0.32\textwidth}
         \centering
         \includegraphics[width=\textwidth]{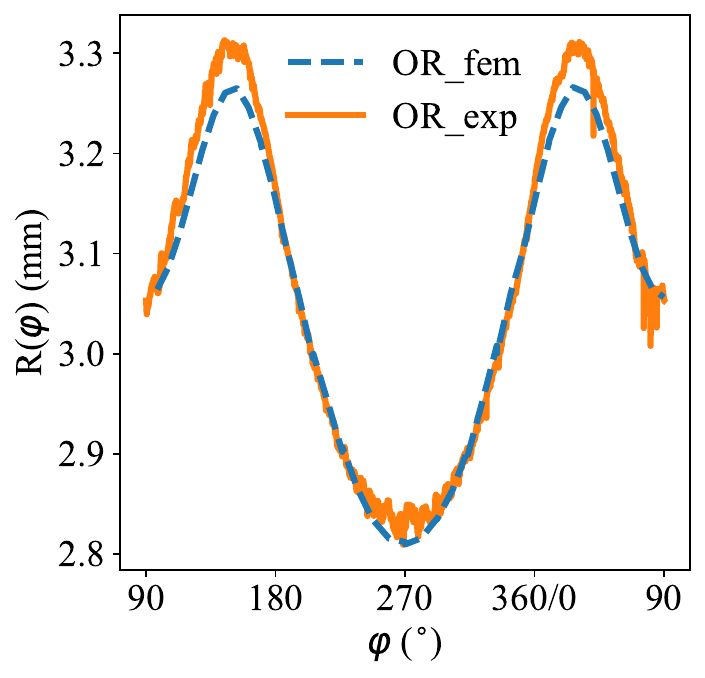}
         \caption{}
         \label{fig:elbow_TD_OR}
     \end{subfigure}
     \caption{Elbow TD cross section: (a) cartesian and (b-c) polar representations for (b) the inner radius IR and (c) the outer radius OR, respectively. Note: to make use of the symmetry along the x-axis, the polar plots start from 90$^{\circ}$ and perform one revolution in a counter-clockwise direction.}
        \label{fig:Elbow_TD_discussion}
\end{figure}

To properly assess the compression bending model developed in this work, Figure \ref{fig:Elbow_TD_discussion} depicts the numerical and experimental curves obtained for the elbow TD cross section, presented in Sections \ref{ssec:Numerical_model} and \ref{ssec:Experimental_campaign}, respectively. A cartesian representation of the distorted geometry is shown in Figure \ref{fig:elbow_TD_fem_exp}, where the difference between the resulting curves is difficult to detect to the naked eye, due to its good agreement. In addition, as both radius (inner and outer) vary along the distorted geometry, a polar representation provides a better option to visualize the ovalization effects. Starting from the inner radius (IR), depicted in Figure \ref{fig:elbow_TD_IR}, the numerical model is able to predict the evolution of IR with a high degree of similitude with respect to the experimental curve, showing two peaks regions connected by a pronounced valley. However, for the outer radius (OR), although the general shape of the curve is obtained, there is a mismatch zone along the top of the peak regions with a maximum value of 39 $\mu$m, as shown in Figure \ref{fig:elbow_TD_OR}. This difference can be explained as the model considers the nominal dimensions of the tube (outer diameter and thickness), therefore, it does not account for any geometrical variability inherent to the tolerances given by the seamless tube manufacturing process. By including this information into the numerical model, it is expected to improve the results.

From the micrographs presented in Section \ref{ssec:Experimental_campaign}, it is worth to mention that the presence of oxide and microfisures detected in the internal walls is a byproduct of the manufacturing process of seamless tubes \cite{Miller2003}. However, as the average flaw size was measured, this information can be used to perform an integrity analysis when designing the positron source target by using Fracture Mechanics criteria.

Regarding the general expression for the yield stress $\sigma_y=\beta_1H_v+\beta_0$, presented at the end of section \ref{ssec:Experimental_campaign}, the slope $\beta_1$ seems to be material independent as the material effects are included in the hardness term $H_v$. Contrary to the negative y-intercept constant $\beta_0$, that presents a material dependency. For example, Tiryakioglu et al., \cite{Tiryakioglu2015b} determined a Vickers hardness-yield stress relationship for the aluminum alloy 7010 in the form $\sigma_y=0.383H_V-182.3$, while the present work proposes a $\beta_0$ = -147.991 for pure tantalum. One possible future use for the obtained expression is to estimate the yield stress on the tantalum tubes of the FCC-ee positron source target prototype after the difussion bonding process of HIPing, so that the possible effect of annealing can be studied.

\section{Conclusions and future works}
\label{sec:Conclusions}

In the present work, the use of compression bending to manufacture a 180$^\circ$ elbow with a pure tantalum seamless tube of dimensions OD 6.35 and ID 4.35 mm was successfully performed. As a result, the following conclusions can be drawn:

\begin{itemize}
    \item The use of 10 mm as the minimum allowable bending radius $R_b$ was confirmed numerically and experimentally. This results goes in line with the reference value of $R_b$ $\geq$ 1.5 OD recommended for rotary draw bending with a multi-ball mandrel.
    \item The developed numerical model is able to predict the deformed shape along the elbow and captures the ovalization of the cross-section. Its use can potentially simplify the manufacturing process for the cooling tubes in other beam intercepting devices as the requirement to fully scan the resulting distorted geometry can be predicted in advance, reducing the design and development phase.
    \item A simplified version of the Khan-Liang (KL) material model for tantalum available in the literature was tested and validated for manufacturing purposes at room temperature. Its use is recommended for the large strain region in further numerical studies at CERN.
    \item The micrographies performed in the cross-section of the tantalum tube before and after plastic bending confirm the integrity of the elbow. The presence of an oxide layer in the internal walls is inherent to the manufacturing process and the presence of small cracks of size 10 $\mu$m can be used to perform integrity studies during the design of the target using fracture mechanics criteria.
    \item An expression to estimate the yield stress based on Vickers hardness  measurements (in MPa) for pure tantalum is proposed:
    \begin{equation*}
        \sigma_y = 0.383H_v-147.991
    \end{equation*}

\end{itemize}

Future work is foreseen to manufacture a prototype of the FCC-ee positron source target with the embedded tantalum cooling tubes. Current R$\&$D efforts are focus on the development of the Hot Isostatic Pressing (HIP) capsule to join by diffusion bonding the tungsten core with the tantalum tubes. Finally, a beam test campaign for the target prototype is planned inside of the PSI Positron Production (P$^3$) experiment. The expected results will be reported in further publications.

\vspace{0.5cm}
\textbf{CRediT authorship contribution statement}:\newline
\textbf{Ramiro Mena-Andrade:} Writing – review \& editing, Writing – original draft, Conceptualization, Methodology, Investigation, Formal analysis, Data curation, Visualization, Validation.
\textbf{Mickaell Crouvizier:} Investigation, Formal analysis.
\textbf{Jean-Philippe Rigaud:} Investigation, Formal analysis.
\textbf{Thibaut Coiffet:} Resources, Methodology.
\textbf{Antonio Perillo-Marcone:} Writing – review \& editing, Supervision, Resources, Project administration, Funding acquisition.

\vspace{0.5cm}
\textbf{Data availability}: Data will be made available on request.

\vspace{0.5cm}
\textbf{Declaration of competing interest}: The authors declare that they have no known competing financial interests or personal relationships that could have appeared to influence the work reported in this paper.

\vspace{0.5cm}
\textbf{Acknowledgements}: This work was done under the auspices of CHART (Swiss Accelerator Research and Technology) Collaboration and the Future Circular Collider Innovation Study (FCCIS). This project has received funding from the European Union's Horizon 2020 research and innovation programme under grant agreement No 951754.

\bibliographystyle{elsarticle-num} 
\bibliography{biblio}


\end{document}